\documentclass[reprint]{revtex4-2}

\usepackage{graphicx}
\usepackage{hyperref}
\usepackage{braket}
\usepackage{cancel}
\usepackage{enumitem}
\usepackage{amssymb, amsmath, amssymb, amsmath}
\usepackage{color}

\begin{document}

\title{Three-Josephson Junctions Flux Qubit Couplings} 

\author{María Hita-Pérez}
\email[]{hitaperezmaria@gmail.com}
\author{Gabriel Jaumà}
\author{Manuel Pino}
\author{Juan José García-Ripoll}

\affiliation{Institute of Fundamental Physics IFF-CSIC, Calle Serrano 113b, Madrid 28006, Spain}

\begin{abstract}
We analyze the coupling of two flux qubits with a general many-body projector into the low-energy subspace. Specifically, we extract the effective Hamiltonians that controls the dynamics of two qubits when they are coupled via a capacitor and/or via a Josephson junction. While the capacitor induces a static charge coupling tunable by design, the Josephson junction produces a magnetic-like interaction easily tunable by replacing the junction with a SQUID. Those two elements allow to engineer qubits Hamiltonians with $XX$, $YY$ and $ZZ$ interactions, including ultra-strongly coupled ones. We present an exhaustive numerical study for two three-Josephson junctions flux qubit that can be directly used in experimental work. The method developed  here, namely the numerical tool to extract qubit effective Hamiltonians at strong coupling, can be applied to replicate our analysis for general systems of many qubits and any type of coupling.
\end{abstract}


\maketitle

Qubit-qubit interactions are a fundamental tool for quantum information processing, both in existing quantum computers\ \cite{nielsen2002quantum} as well as in quantum simulators\ \cite{buluta2009quantum,cirac2012goals,ballester2012quantum}. Qubits that interact directly by physical means, usually develop dipolar-like couplings with a well-defined orientation---$\sigma^x_1\sigma^x_2$ or $\sigma^z_1\sigma^z_2$ on the qubit basis---due to the electromagnetic nature of those interactions. This is the case of transmons and flux qubits, objects which exhibit capacitive and inductive type couplings, respectively.

In recent years there has been growing interest in enlarging the families of interactions between superconducting qubits\ \cite{collodo2020,ozfidan2020,Consani2020,Kerman2019} while preserving their strength. This has obvious advantages in the world of quantum computing by making possible the implementation of a rich family of gates\ \cite{arute2019quantum}. It is also relevant in the development of quantum simulators, enabling non-stoquastic models that are harder to simulate classically\ \cite{bravyi2006,susa2017,albash2019role,hormozi2017, ozfidan2020}, and which could eventually lead to universal adiabatic quantum computation\ \cite{albash2018adiabatic,kempe2006complexity,oliveira2005complexity}.In this context, flux qubits\ \cite{ozfidan2020,Consani2020,Kerman2019} gain extra relevance when compared to transmon qubits \ \cite{collodo2020}. The large anharmonicity of flux qubits ensures the preservation of a well-defined qubit basis even for ultra-strong couplings that dominate over the qubit's local Hamiltonian. In comparison, an ultra-strong interaction between transmon qubits enables transitions from qubit states to the weakly anharmonic excitations, so that the effective dynamics cannot be captured by a spin-$1/2$ model.

In this work we perform an exhaustive study of inductive and capacitive couplings between three-Josephson-junction flux qubits (3JJQ)\ \cite{orlando}, as an alternative to the commonly used rf-SQUID qubits \ \cite{ozfidan2020}, analyzing the origin of the interactions and the effective Hamiltonians that they produce. The capacitive interaction is implemented via a capacitor joining the two qubits, and the inductive-like coupling comes from a shared Josephson junction. One of our main goals is to determine the best circuit designs and parameter region that provides large couplings while retaining acceptable qubits properties. This is relevant in a scenario where there is a non-linear dependence of physical properties on the circuit parameters.

We are interested in the extraction of the effective low-energy Hamiltonian for this type of interacting circuits. Our tool to do so is an improved version of the Schrieffer-Wolff transformation (SWT) introduced in  Ref.~\onlinecite{Consani2020}. This method has allowed us to demonstrate that capacitively and inductively coupled 3JJQs reproduce a fairly large family of spin-1/2 models, with strong and non-stoquastic couplings on the qubit basis. If we denote the Hamiltonian of one flux qubit as $H_q= \frac{\Delta}{2} \sigma^z$, the Josephson junction coupling produces a magnetic-like interaction of the $J_{xx}\sigma^x_1\sigma^x_2$ form, while the capacitive coupling creates interactions along orthogonal directions $J_{yy}\sigma_1^y\sigma_2^y$ and $J_{zz}\sigma_1^z\sigma_2^z$\ \cite{satoh2015ising,ozfidan2020,Consani2020,hitaperez2021ultrastrong}. Our study shows regimes of strong coupling with $J_{xx}\approx J_{yy}\approx J_{zz}> \Delta$ using realistic qubit designs as those proposed by Ref.~\onlinecite{Chiorescu_2003}.\\

We first analyze the numerical method that allows us to extract the effective Hamiltonians $H_\text{eff}$ from a general many-body system by means of the SWT. A unitary mapping between two non-orthogonal subspaces of the same dimension \ \cite{schrieffer1966relation,bravyi2011}, represented by projectors  $P_0$ and $P$
\begin{equation}
    U = \sqrt{(2P_0 - \openone_0)(2P - \openone) },
    \label{e. def SWT}
\end{equation}
satisfying $UP U^\dagger = P_0$. Here, $P_0$ projects onto the low-energy or qubit subspace of a set of non-interacting circuits $H_0 = \sum_i H_i$, {being Hi the independent Hamiltonian of each of them, and $P$ accesses the low-energy subspace when those circuits interact producing a Hamiltonian $H=H_0+V$. $\openone_0$ and $\openone$ are the projectors onto the full set of eigenstates of the Hamiltonians. For moderate interactions the subspace $P$ remains gapped from non-qubit states and allows us to interpret the physical interaction $V$ as a coupling between qubits in the unperturbed qubit subspace}
\begin{equation}
H_\text{eff}= P_0UPHPU^\dagger P_0. \label{e. H_eff}
\end{equation}

Consani \textit{et al.}\ \cite{Consani2020} computed $U$ directly from its definition \eqref{e. def SWT}, and showed that $H_\text{eff}$ includes qubit-qubit interactions\ \cite{Consani2020} not captured by perturbative methods. To make the method affordable, Consani \textit{et al.}\ \cite{Consani2020} express $H$ in the basis of eigenstates of the uncoupled qubits with up to $N_T$ states---a number determined by converegence---. In this basis, $H$ is approximately diagonalized to recover the interacting eigenstates and $P$, and $U$ is computed using equation \eqref{e. def SWT}. This step dominates the complexity of the algorithm, due to working with matrices of size $N_T\times N_T$.

We propose to compute the rank-$d$ matrices $P_0U P$ and $PU^\dagger P_0$, using only the $d$ eigenstates of $H_0$ and $H$ that span their low-energy subspaces. Thus, instead of computing $U$ in the full basis, we only need to estimate $d \times d$ matrices. The cost of the algorithm is now dominated by the calculation of the $d$ eigenstates, a step also present in Consani's work, which is done efficiently using Lanczos techniques. To develop this simplification, we note that $P_0UP$ involves a rank-$d$ transformation $A$
\begin{equation}
  P_0UP = \sum_{i,j=1}^d A_{ij} \ket{\psi_i^0}\!\bra{\psi_j}  =P_0AP,
  \label{e. Anm}
\end{equation}
among the basis $\ket{\psi_j^0}$ and $\ket{\psi_j}$ in which we represent $P_0$ and $P$. Using these basis, $H_\text{eff}$ becomes
\begin{equation}
  H_\text{eff} = \sum_{ijk=1\ldots d}\ket{\psi_i^0}\!A_{ij}\braket{\psi_j|H|\psi_j}(A^\dagger)_{jk}\bra{\psi^0_k}.
  \label{eq:effective-H}
\end{equation}
We introduce a rank-$d$ operator $B=P_0P$, such that $(P_0UP)^2 =(P_0AP)^2 = P_0AB^\dagger AP= P_0BP$, hence $AB^\dagger A=B$. And its singular value decomposition\ \cite{golub1996matrix} $B=W\Sigma V^{\dagger}$, in terms of two unitary transformations $W,V \in \mathbb{C}^{d\times d}$, and a non-negative diagonal matrix $\Sigma$. Using these matrices, we can verify that the previous equation is satisfied by $A  = W V^\dagger$.\\

\begin{figure}
\centering
  \includegraphics[width=0.48\textwidth]{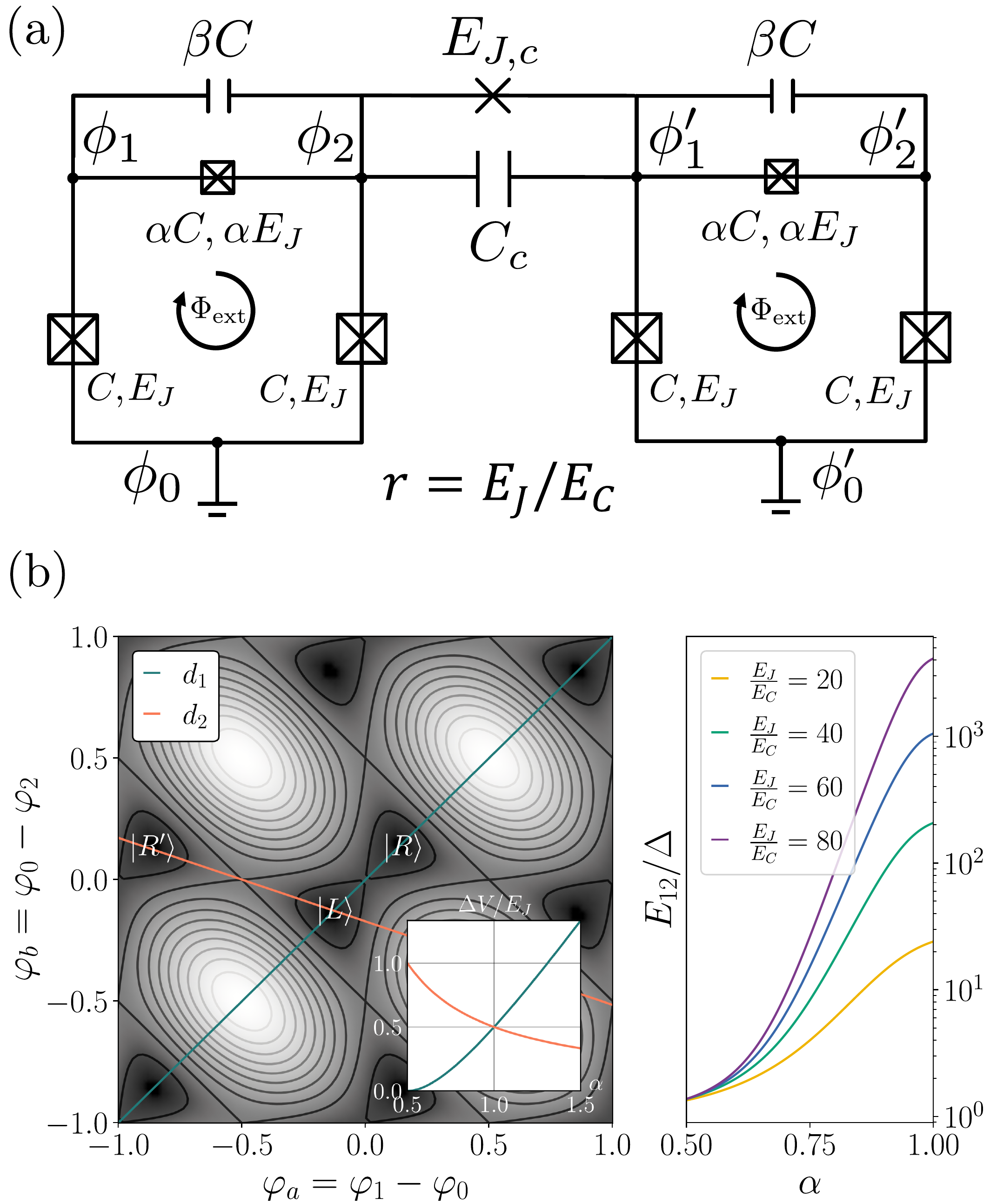}
  \caption{ (a) Two identical c-shunted 3JJQs with grounds in nodes $\phi_0$ and $\phi_0'$ coupled through a capacitor with capacitance $C_c$ and a Josephson junction with Josephson energy $E_{J,c}$.
  (b) Analysis of one single 3JJQ. At left panel, unit cell for the periodic nonlinear potential  at full frustration, the potential strength is represented in a scale of colors from the maximum in white to the minimum in black, as a function of the phase variables $\varphi_i=2\pi\phi_i/\Phi_0$. The two qubit states are indicated as $\ket{L},\ket{R}$ for the unit cell ($\ket{R'}$ correspond to a qubit state at an adjacent cell). The inset shows the barrier height through the intra and inter-cell tunneling directions. Right panel: ratio of energy differences between $E_{12}/\Delta$, with $E_{12}$ the  energy difference between second and first exited levels and $\Delta$ the qubit gap, as a function of $\alpha$ for multiple values of $r=E_J/E_C$ and $\beta=0$. Notice that the relative anharmonicity is $\alpha_r =E_{12}/\Delta-1$.}
  \label{f. potencial y anh}
\end{figure}

We are interested on using the SWT on pairs of superconducting qubits, with a bare Hamiltonian $H_0=H_1+H_2$ and an interaction $\gamma V$ controlled by a parameter $\gamma$. Provided there exists a gapped low-energy subspace $P$, we can express the circuit model $H=H_0+\gamma V$ in the charge basis and compute $H_\text{eff}$, which we expand in the complete basis of Pauli matrices as
\begin{equation}
  H_\text{eff} = \sum_{i=x,y,z}\left(\frac{h_{1i}}{2}\sigma_1^i + \frac{h_{2i}}{2}\sigma_2^i\right) + \sum_{i,j=x,y,x} J_{ij}\sigma_1^i\sigma_2^j.\label{e. Final Hamiltonian}
\end{equation} \par 

We apply this technique to  3JJQs\ \cite{orlando, orlando2}, which are composed of a superconducting loop interrupted by three Josephson junctions, two identical and one a factor $\alpha$ times smaller [cf. Fig.~\ref{f. potencial y anh}]. At full frustration $\Phi_{\rm ext} = \frac{1}{2}\Phi_0$, the periodic inductive potential has a unit cell depicted in Fig.~\ref{f. potencial y anh}(b). For $\alpha> 0.5$ this unit cell has two minima that correspond to two persistent current states flowing in opposite direction. Quantum tunneling couples these current states, creating an effective qubit subspace $H_q=\frac{\Delta}{2}\sigma^z$ with a gap $\Delta$. We choose our qubit parameters in the range $0.6<\alpha < 0.9$. The first inequality guarantees that the qubit subspace $P_0$ is gaped from non-qubit states with a large anharmonicity $E_{21}>2\Delta$. The second inequality ensures that the intra-cell tunneling along direction $d_1$ dominates over the inter-cell tunneling $d_2$ [cf. Fig.~\ref{f. potencial y anh}(b)], reducing the sensitivity to phase slips and charge noise\ \cite{orlando}. We also study qubits with a shunting capacitor in parallel to the small junction because, despite their reduced anharmonicity, they exhibit lower sensitivity to flux noise and better reproducibility\ \cite{yan2016}.

\begin{figure*}
    \centering \includegraphics[width=\linewidth]{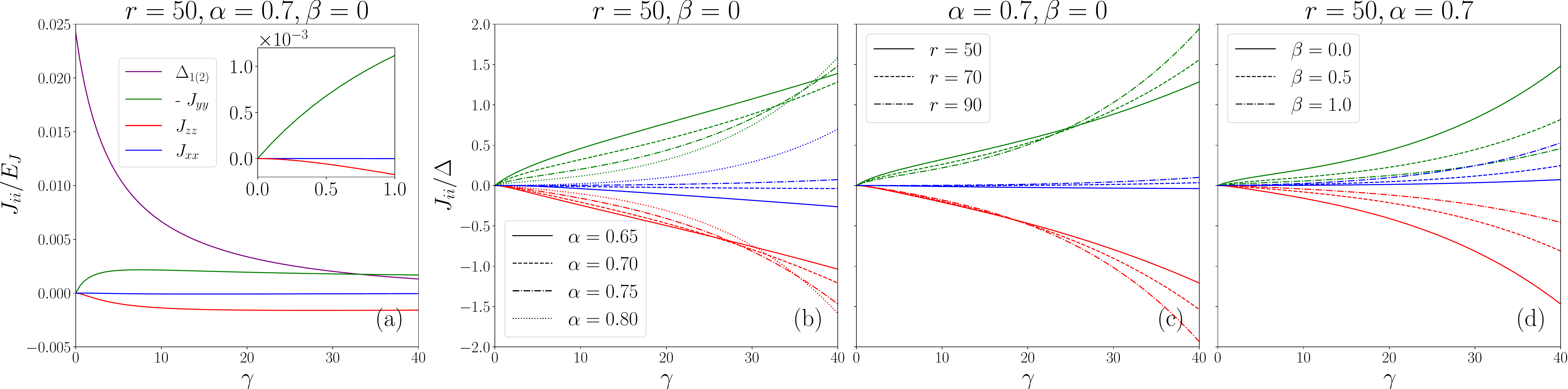}
    \caption{Coupling strengths for two 3JJQs with ground in $\phi_0$($\phi_0'$) coupled through a capacitor connecting nodes $\phi_2-\phi_1'$. (a) Effective Hamiltonian parameters as a function of $\gamma$ for $\alpha=0.70$, $r=50$ and $\beta=0$. (b-d) Ratios between the coupling strengths ($J_{ii}$) and the qubit gap ($\Delta$) for fixed: (b) $r$ and $\beta$,  (c) $\alpha$ and $\beta$, (d) $\alpha$ and $r$. The legend in plot (a) holds for (b-d). We represent $-J_{yy}$ for the sake of clarity.}
    \label{f. capacitive coupling}
\end{figure*}

Now we move on the discussion of qubit's coupling. We consider circuits such as the one in Fig.~\ref{f. potencial y anh}(a), as well as 11 other topologies that change the places where the capacitor and inductor are attached and the positions of the grounds. These circuits are quantized and brought to the form $H=H_1+ H_2 +H_\text{int}$, where the $H_{1,2}$ are identified as renormalized single-qubit Hamiltonians, and the remaining interactions are grouped into a perturbation $H_\text{int}:=\gamma V$ (see Supplementary Material A). The complete Hamiltonian is then analyzed using the SWT and brought into a form~\eqref{e. Final Hamiltonian}. Working at the symmetry point, $\Phi_{\rm ext} = \frac{1}{2}\Phi_0$, we find that the only nonzero terms are the renormalized single-qubit gaps $h_{1z}=\Delta_1$ and $h_{2z}=\Delta_2$, and the qubit-qubit interactions $\sigma^x_1\sigma^x_2$, $\sigma^y_1\sigma^y_2$ and $\sigma^z_1\sigma^z_2$, modulated by the coupling strengths $J_{xx}$, $J_{yy}$ and $J_{zz}$.\\

Let us discuss first the capacitive coupling of two identical 3JJQs shown in Fig.~\ref{f. potencial y anh}(a) with $C_c=\gamma C$, $E_{J,c}=0$ and grounds at $\phi_0=\phi_0'=0$. From the analytical perturbative study performed in Ref.~\onlinecite{hitaperez2021ultrastrong}, we expect three types of interactions $\sigma^y_1\sigma^y_2,\ \sigma^z_1\sigma^z_2$ and $\sigma^x_1\sigma^x_2$, corresponding to first, second and third order coupling in $\gamma$. The $J_{yy}$ terms are explained by the matrix elements of the charge operator within the qubit subspace\ \cite{hitaperez2021ultrastrong}, while $J_{zz}$ and $J_{xx}$ are interactions mediated by states outside the qubit space. Perturbation theory agrees qualitatively with the numerically exact results shown in Fig.~\ref{f. capacitive coupling}(a) for small interaction strengths $\gamma$, but its prediction fails for moderate interactions where the $J_{yy}$ and $J_{zz}$ couplings reach a maximum similar in magnitude and then slowly start to decay.

Figs.~\ref{f. capacitive coupling}(b-d) display the growth of the relative interaction strength $J/\Delta$ for the design parameters in Fig.~\ref{f. potencial y anh}(a) ($\alpha$, $r$ and $\beta$), illustrating the crossover from weak $J/\Delta\ll 1$ to strong coupling regime $J/\Delta\approx 1$. For small $\gamma$, the behavior of the coupling is dominated by the perturbative tendencies in $J_{ii}$. For larger couplings, the growth of $J/\Delta$ is dominated by the exponential decrease \ \citep{orlando} of the gap $\log{(\Delta)}= \mathcal{O} (\sqrt{C_q/C})$  with the renormalized qubit capacitance, which grows with $\gamma,\beta$ and $\alpha$. This competition explains the non-monotonical behaviour found in $J_{zz}/\Delta,J_{yy}/\Delta$ with respect to $\alpha,r$ [cf. Figs.~\ref{f. capacitive coupling}(b-c)], as $J$ decreases while $1/\Delta$ increases with those parameters. Finally, for the limited range of $\gamma$ where the gap is not negligible, $J_{zz}/\Delta$ and $J_{yy}/\Delta$ always decrease with the shunting $\beta$.

Note that at the same time as the intra-cell tunneling is suppressed, which produces the exponential decay in the qubit's gap, the tunneling along the $d_2$ direction may get activated, see Fig.\ \ref{f. potencial y anh}. This phenomenon is due to the  renormalization of the capacitances along different directions and is thus dependent on the qubit's parameters and the circuit topology. A consequence of this activation is the fast growth of the $J_{xx}\sigma^x_1\sigma^x_2$ interaction. This is, in our opinion, a regime to be avoided. First, because the $J_{xx}$ can be obtained by other (inductive) means. And second, because the activation of the inter-cell tunneling is accompanied by a greater sensitivity to electrostatic field fluctuations.

Different coupling topologies produce qualitatively similar plots, although the relative coupling strength $J_{ii}/\Delta=1$ may be reached for lower or higher values of the capacitance $\gamma$, and the relative sign of the interactions might change. We have also studied different grounding schemes. Topologically, there are two distinct combinations: we can place the grounds between the small and big junctions ---\textit{i.e.} $\phi_1=0$ or $\phi_{2}=0$---or we can place them between the big junctions $\phi_0=\phi_{0'}=0$. Choosing between $\phi_1=0$ or $\phi_2=0$ is equivalent to flipping the flux passing through the qubit, and changes the sign of the $\sigma^y$ and $\sigma^x$ operators. If we choose topologically equivalent grounds for both qubits, we obtain coupling strengths with similar magnitude as the ones seen before. However, there are somewhat pathological choices---e.g. $\phi_0=\phi_{1'}=0$ connecting nodes 0 and $1'$---where the qubits experience different renormalizations and their gaps differ as interaction grows.

It must be remarked that for all choices of connecting nodes and ground nodes we always obtain \emph{both} $J_{yy}\sigma^y_1\sigma^y_2$ and $J_{zz}\sigma^z_1\sigma^z_2$ interactions simultaneously, with very similar magnitude. This means that we can engineer effective qubit-qubit interactions of the approximate form $J(\sigma_1^z\sigma_2^z\pm\sigma_1^y\sigma_2^y)$, with  $J\approx \Delta$ and with a sign that depends on the topology. This could produce a spectral signature that is similar to the one observed in Ref.~\onlinecite{ozfidan2020}, but without the flexibility suggested in that experimental work.\\

We have additionally studied the inductive coupling  between two identical 3JJQs  with a Josephson junction, using the circuit topology in Fig.~\ref{f. potencial y anh}(a), with $E_{J,c}=\gamma E_{J}$ and $C_c=0$, but with grounds $\phi_1=\phi_{2'}=0$ (notice that we neglect the junction's capacitance). The interaction is so strong that around $\gamma\approx 0.1$ it produces a full hybridization of the low and high energy subspaces where we cannot isolate a qubit subspace.

\begin{figure}
    \centering
    \includegraphics[width=\linewidth]{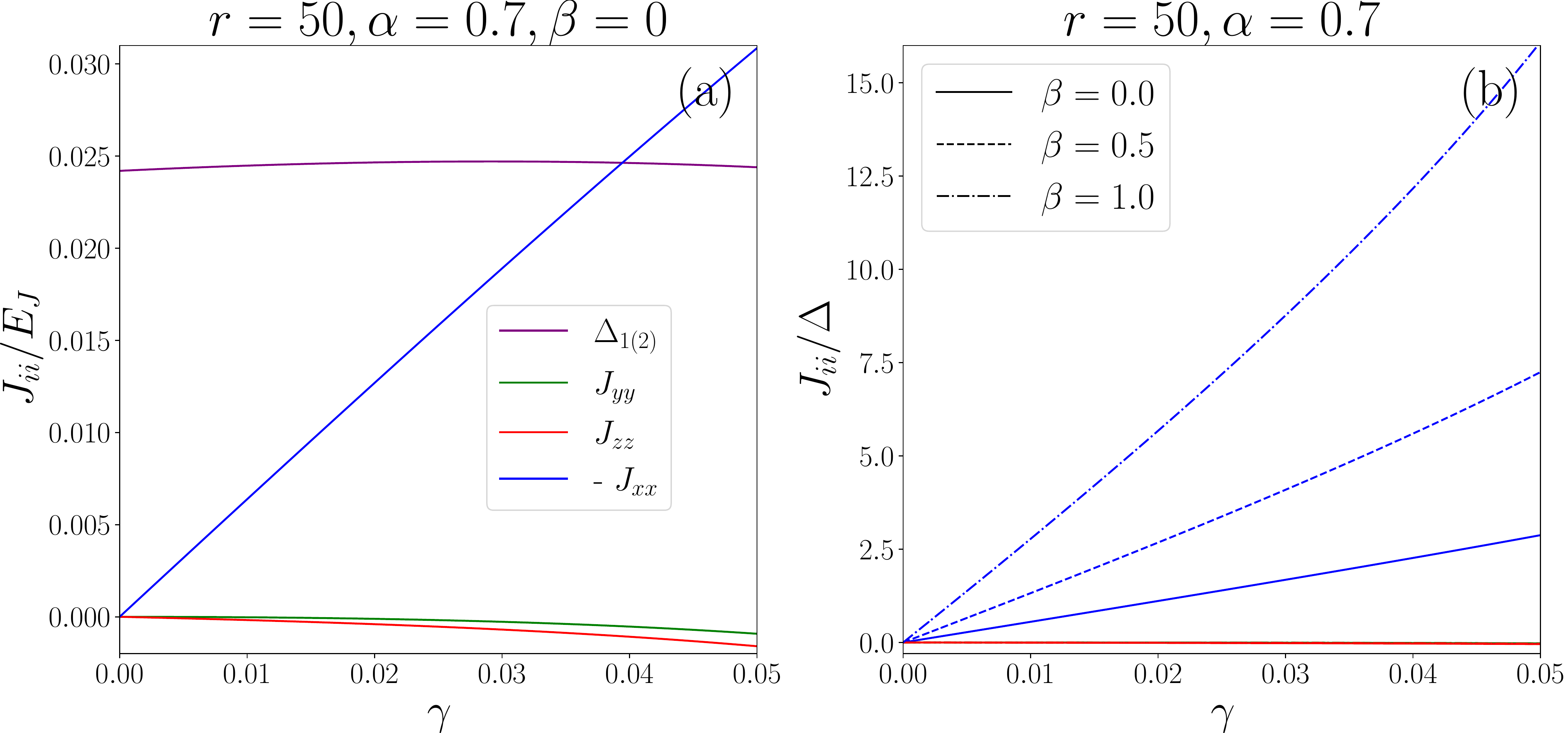}
    \caption{Coupling strengths for two 3JJQs with ground in $\phi_1$($\phi_2'$) coupled through a Josephson junction connecting nodes $\phi_2-\phi_1'$. (a) Effective Hamiltonian parameters as a function of $\gamma$ for $\alpha=0.70$, $r=50$ and $\beta=0$. (b) Ratios between the coupling strengths ($J_{ii}$) and the qubit gap ($\Delta$) for fixed $\alpha$ and $r$. We represent $-J_{xx}$ for the sake of clarity.}
    \label{f. inductive coupling}
\end{figure}

Before this regime, for $0<\gamma<0.05$, as illustrated by Fig.~\ref{f. inductive coupling}(a), interactions are dominated by the coupling $J_{xx}\sigma_1^x\sigma_2^x$ between the effective dipolar magnetic moments of both qubits. In addition to this, we find some residual $J_{zz}\sigma^z_1\sigma^z_2$ and $J_{yy}\sigma^y_1\sigma^y_2$ contributions, that are up to three orders of magnitude weaker and can be neglected.

The dependency of the coupling strengths on the 3JJQ parameters offers a simple picture, where the dominant inductive term $J_{xx}\sigma^x_1\sigma^x_2$ grows with $\alpha$ and $r$ (data not shown). This tendency is accompanied by a reduction of the qubit gap for increasing $\alpha$ and $r$. Finally, as it can be extracted from Fig.~\ref{f. inductive coupling}(b), adding a shunting capacitor to the 3JJQs reduces the qubits gap while strengthening the $\sigma^x_2\sigma^x_2$ inductive coupling. This allows for arbitrarily large ratios between the coupling strength and the gap of the qubit leading to ultrastrong coupling but also favors the crossing between levels inside and outside the qubit subspace for increasingly small values of $\gamma$.

Similar to the capacitive circuit, changing the circuit topology does not affect the qualitative behavior of the interaction with the coupling strength $\gamma$. At most, the choice of coupling and ground nodes can speed up or slow down the growth of interactions with $\gamma$ or change the sign of the corresponding qubit operator.\\

We have used dimensionless quantities $(\alpha,\beta,\gamma,r)$ in our previous analysis. We now present results using dimensional parameters close to the experimental ones in Ref.~\onlinecite{Chiorescu_2003}. Employing the machinery previously developed, we show that those parameters allow obtaining qubit-qubit strong couplings that can be implemented in experiments.

\begin{figure}
    \centering
    \includegraphics[width=\linewidth]{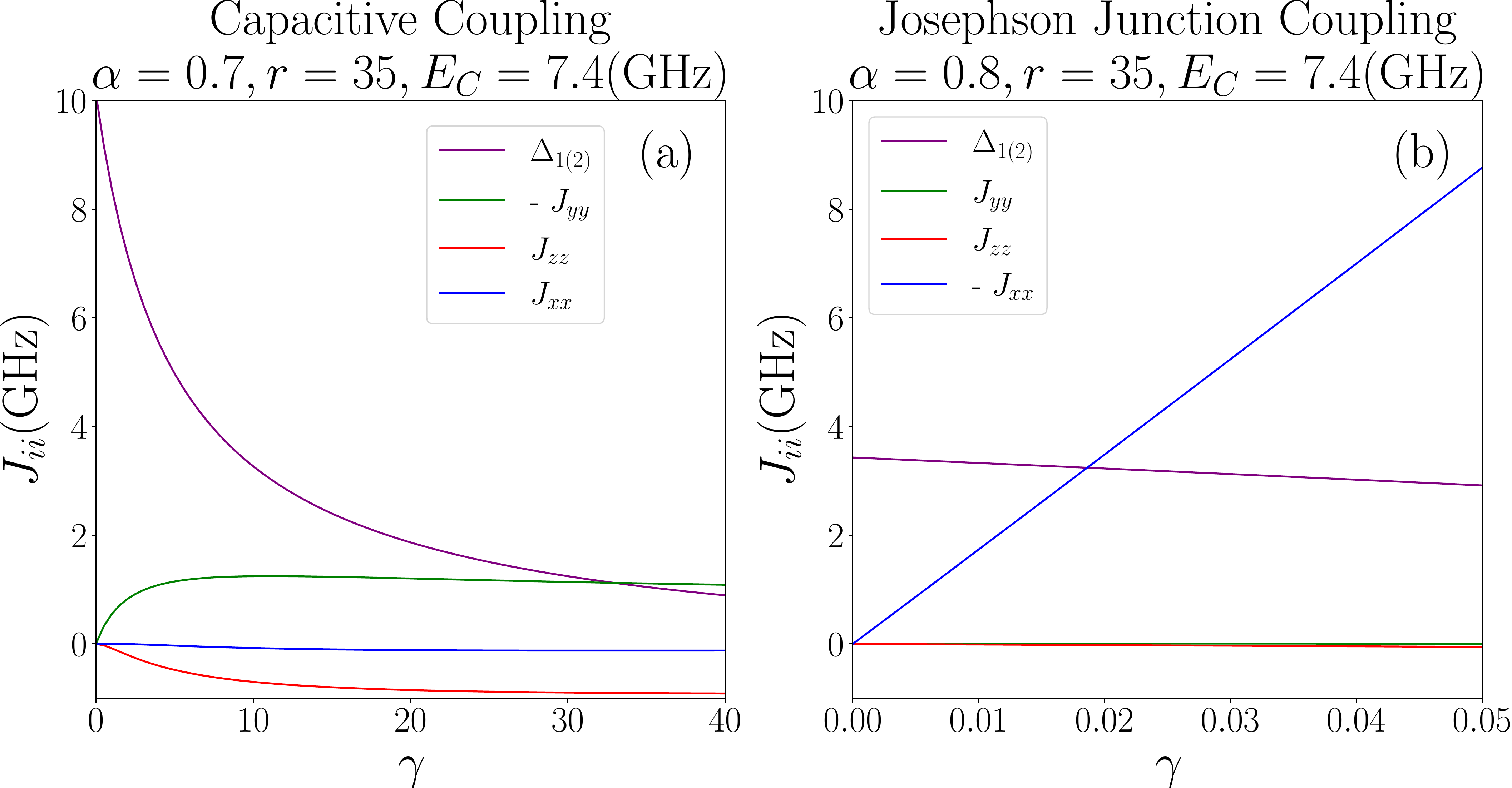}
    \caption{Coupling strengths for the reference circuits using real experimental 3JJQs parameters \ \protect\cite{Chiorescu_2003}.(a) Effective Hamiltonian parameters as a function of $\gamma$ for $\alpha=0.70$, $r=35$, $E_C=7.4$ (GHz) and $\beta=0$ for the capacitive coupling. (b) Effective Hamiltonian parameters as a function of $\gamma$ for $\alpha=0.80$, $r=35$, $E_C=7.4$ (GHz) and $\beta=0$ for the Josephson junction coupling (including the capacitor that accompanies it). We represent $-J_{yy}$ in (a) and $-J_{xx}$ in (b) for the sake of clarity.}
    \label{f. realistic coupling}
\end{figure}

For the capacitive coupling, the parameters used in Ref.~\onlinecite{Chiorescu_2003} are ideal to produce a Hamiltonian of the form~\eqref{e. Final Hamiltonian} with a negligible magnetic interaction. Indeed, the small ratio between the Josephson energy and the capacitive energy ($E_C=7.4$), $r=35$, ensures that $J_{xx}\approx 0$ for the circuit configuration considered above, while $\alpha=0.8$ ensures a good separation of the two-qubit subspace from higher-energy excitations, even after the coupling. However, such a large value of $\alpha$ gives rise to a strong renormalization of the gap, limiting the coupling constants to $J_{ii}<0.2$ GHz, a small value for practical applications.

To overcome this limit, we propose to slightly reduce the qubit's anharmonicity, using $\alpha=0.7$. As seen in Fig.~\ref{f. realistic coupling}(a), this modification produces effective qubit couplings  $J(\sigma_1^y\sigma_2^y+\sigma_1^z\sigma_2^z)$ which are strong ($J/\Delta\approx1$) and values of the qubit gap and interactions that can be measured experimentally, free of thermal fluctuations. Note that models with $J(\sigma_1^y\sigma_2^y-\sigma_1^z\sigma_2^z)$ couplings are obtained by changing one of the flux threading one qubit, or the connection topology.

Fig.~\ref{f. realistic coupling}(b) presents a similar analysis for the inductive coupling, using Chiorescu et al's original qubits\ \cite{Chiorescu_2003}, considering both the junction and its capacitance. Since the inductive coupling between qubits is extremely strong and does not allow for large values of $\gamma$, it can be seen that the added capacitor does not really affect the coupling results. Note the lack of strong qubit renormalization allows for strong $\sigma^x_1\sigma^x_2$ coupling ($J_{xx}/\Delta>2$) without having to modify the original qubit.\\

Summing up, we have performed an extensive analysis of a system composed of two 3JJQs coupled via a capacitor and a Josephson junction. This type of circuits can be implemented experimentally to produce qubits models with strong coupling in different directions. Indeed, we have obtained arbitrary interactions of the form $J^\text{cap}(\sigma_1^y\sigma_2^y\pm\sigma_1^z\sigma_2^z)+(J_{xx}^\text{cap}+J_{xx}^\text{JJ})\sigma_1^x\sigma_2^x$.

Our results confirm the idea that flux qubits may be used to simulate strong non-stoquastic spin Hamiltonians, but also reveal that not all interactions are independent as found by the simultaneous appearance of $\sigma^y_1\sigma^y_2$ and $\sigma^z_1\sigma^z_2$ terms. This may have consequences for the interpretation of works that argue the classical simulability of superconducting quantum circuits\ \cite{ciani2020,halverson2020efficient}.

Out of the two interactions studied, $J_{xx}^\text{JJ}$ admits straightforward tuneability, replacing the junction with a dc-SQUID. We believe that this avoids the complex dynamic of the usual rf-SQUID tuneable couplers, lifting its geometric constrains\ \cite{Consani2020,ozfidan2020,kafri,Weber_2017,allman2010_a}. We also believe that the capacitive interactions $J^\text{cap}$ can be tuned with the help of mediating circuits, such as capacitively connected, tuneable frequency qubits and resonators.

Designing the type of circuits discussed here is a difficult task. A source of problems is the non-linear dependence of physical quantities on the design parameters, as the exponential renormalization of the qubit gap with the coupling capacitance. Our methods could help experimentalists in the design and optimization of qubits and couplers. As an example of the utility of our work, we have found acceptable parameters close to earlier experimental setups, which provide strong capacitively or inductively coupled qubits.

Additionally, we have developed an improved numerical scheme to derive effective Hamiltonians of coupled quantum systems, reducing the complexity of the original algorithm in Ref.~\onlinecite{Consani2020} down to the size of the computational space under analysis, times the size for the representation of the relevant eigenstates. This allows us to treat a wider class of models such as qubits interacting with resonators\ \cite{hitaperez2021ultrastrong}, and can be extended to treat larger systems if a clever representation---e.g. tensor networks---is used to describe the low-energy states. This way, we expect to scale our simulations up to $4$ or $5$ qubits, exploring the gap renormalization due to capacitive couplings, and whether the effective models can still be obtained as a sum of two- or at most three-body operators in the case of strong qubit interactions.

Finally, our work leaves open questions, such as the application of our coupling scheme in the context of quantum computation, where the tunability of the capacitive couplings may become relevant. We expect to analyze this question in future works, inducing a mediated capacitive coupling via, for instance, microwave resonators\ \cite{hitaperez2021ultrastrong} or other qubits\ \cite{arute2019quantum}.

\section*{Supplementary Material}

See supplementary material for the full derivation of the system's Hamiltonian (Supplementary Material A) and a further study of the multiple circuit configurations (Supplementary Material B).

\begin{acknowledgments}

This work has been supported by European Commission FET-Open project AVaQus GA 899561 and CSIC
Quantum Technologies Platform PTI-001. Financial support by Fundación General CISC (Programa Comfuturo)
is acknowledged. The numerical computations have been
performed in the cluster Trueno of the CSIC.
\end{acknowledgments}

\section*{Data Availability Statement}

The data that support the findings of this study are available from the corresponding author upon reasonable request.

\section*{Authors Declarations}
The authors have no conflicts to disclose.\\


\bibliography{bibliography.bib}

\begin{thebibliography}{31}%
\makeatletter
\providecommand \@ifxundefined [1]{%
 \@ifx{#1\undefined}
}%
\providecommand \@ifnum [1]{%
 \ifnum #1\expandafter \@firstoftwo
 \else \expandafter \@secondoftwo
 \fi
}%
\providecommand \@ifx [1]{%
 \ifx #1\expandafter \@firstoftwo
 \else \expandafter \@secondoftwo
 \fi
}%
\providecommand \natexlab [1]{#1}%
\providecommand \enquote  [1]{``#1''}%
\providecommand \bibnamefont  [1]{#1}%
\providecommand \bibfnamefont [1]{#1}%
\providecommand \citenamefont [1]{#1}%
\providecommand \href@noop [0]{\@secondoftwo}%
\providecommand \href [0]{\begingroup \@sanitize@url \@href}%
\providecommand \@href[1]{\@@startlink{#1}\@@href}%
\providecommand \@@href[1]{\endgroup#1\@@endlink}%
\providecommand \@sanitize@url [0]{\catcode `\\12\catcode `\$12\catcode
  `\&12\catcode `\#12\catcode `\^12\catcode `\_12\catcode `\%12\relax}%
\providecommand \@@startlink[1]{}%
\providecommand \@@endlink[0]{}%
\providecommand \url  [0]{\begingroup\@sanitize@url \@url }%
\providecommand \@url [1]{\endgroup\@href {#1}{\urlprefix }}%
\providecommand \urlprefix  [0]{URL }%
\providecommand \Eprint [0]{\href }%
\providecommand \doibase [0]{https://doi.org/}%
\providecommand \selectlanguage [0]{\@gobble}%
\providecommand \bibinfo  [0]{\@secondoftwo}%
\providecommand \bibfield  [0]{\@secondoftwo}%
\providecommand \translation [1]{[#1]}%
\providecommand \BibitemOpen [0]{}%
\providecommand \bibitemStop [0]{}%
\providecommand \bibitemNoStop [0]{.\EOS\space}%
\providecommand \EOS [0]{\spacefactor3000\relax}%
\providecommand \BibitemShut  [1]{\csname bibitem#1\endcsname}%
\let\auto@bib@innerbib\@empty
\bibitem [{\citenamefont {Nielsen}\ and\ \citenamefont
  {Chuang}(2010)}]{nielsen2002quantum}%
  \BibitemOpen
  \bibfield  {author} {\bibinfo {author} {\bibfnamefont {M.~A.}\ \bibnamefont
  {Nielsen}}\ and\ \bibinfo {author} {\bibfnamefont {I.~L.}\ \bibnamefont
  {Chuang}},\ }\href {https://doi.org/10.1017/CBO9780511976667} {\emph
  {\bibinfo {title} {Quantum Computation and Quantum Information: 10th
  Anniversary Edition}}}\ (\bibinfo  {publisher} {Cambridge University Press},\
  \bibinfo {year} {2010})\BibitemShut {NoStop}%
\bibitem [{\citenamefont {Buluta}\ and\ \citenamefont
  {Nori}(2009)}]{buluta2009quantum}%
  \BibitemOpen
  \bibfield  {author} {\bibinfo {author} {\bibfnamefont {I.}~\bibnamefont
  {Buluta}}\ and\ \bibinfo {author} {\bibfnamefont {F.}~\bibnamefont {Nori}},\
  }\bibfield  {title} {\bibinfo {title} {Quantum simulators},\ }\href
  {https://doi.org/10.1126/science.1177838} {\bibfield  {journal} {\bibinfo
  {journal} {Science}\ }\textbf {\bibinfo {volume} {326}},\ \bibinfo {pages}
  {108} (\bibinfo {year} {2009})}\BibitemShut {NoStop}%
\bibitem [{\citenamefont {Cirac}\ and\ \citenamefont
  {Zoller}(2012)}]{cirac2012goals}%
  \BibitemOpen
  \bibfield  {author} {\bibinfo {author} {\bibfnamefont {J.}~\bibnamefont
  {Cirac}}\ and\ \bibinfo {author} {\bibfnamefont {P.}~\bibnamefont {Zoller}},\
  }\bibfield  {title} {\bibinfo {title} {Goals and opportunities in quantum
  simulation},\ }\href {https://doi.org/10.1038/nphys2275} {\bibfield
  {journal} {\bibinfo  {journal} {Nature Physics}\ }\textbf {\bibinfo {volume}
  {8}},\ \bibinfo {pages} {264} (\bibinfo {year} {2012})}\BibitemShut {NoStop}%
\bibitem [{\citenamefont {Ballester}\ \emph {et~al.}(2012)\citenamefont
  {Ballester}, \citenamefont {Romero}, \citenamefont {Garc{\'\i}a-Ripoll},
  \citenamefont {Deppe},\ and\ \citenamefont {Solano}}]{ballester2012quantum}%
  \BibitemOpen
  \bibfield  {author} {\bibinfo {author} {\bibfnamefont {D.}~\bibnamefont
  {Ballester}}, \bibinfo {author} {\bibfnamefont {G.}~\bibnamefont {Romero}},
  \bibinfo {author} {\bibfnamefont {J.~J.}\ \bibnamefont {Garc{\'\i}a-Ripoll}},
  \bibinfo {author} {\bibfnamefont {F.}~\bibnamefont {Deppe}},\ and\ \bibinfo
  {author} {\bibfnamefont {E.}~\bibnamefont {Solano}},\ }\bibfield  {title}
  {\bibinfo {title} {Quantum simulation of the ultrastrong-coupling dynamics in
  circuit quantum electrodynamics},\ }\href
  {https://doi.org/10.1103/physrevx.2.021007} {\bibfield  {journal} {\bibinfo
  {journal} {Physical Review X}\ }\textbf {\bibinfo {volume} {2}},\ \bibinfo
  {pages} {021007} (\bibinfo {year} {2012})}\BibitemShut {NoStop}%
\bibitem [{\citenamefont {Collodo}\ \emph {et~al.}(2020)\citenamefont
  {Collodo}, \citenamefont {Herrmann}, \citenamefont {Lacroix}, \citenamefont
  {Andersen}, \citenamefont {Remm}, \citenamefont {Lazar}, \citenamefont
  {Besse}, \citenamefont {Walter}, \citenamefont {Wallraff},\ and\
  \citenamefont {Eichler}}]{collodo2020}%
  \BibitemOpen
  \bibfield  {author} {\bibinfo {author} {\bibfnamefont {M.~C.}\ \bibnamefont
  {Collodo}}, \bibinfo {author} {\bibfnamefont {J.}~\bibnamefont {Herrmann}},
  \bibinfo {author} {\bibfnamefont {N.}~\bibnamefont {Lacroix}}, \bibinfo
  {author} {\bibfnamefont {C.~K.}\ \bibnamefont {Andersen}}, \bibinfo {author}
  {\bibfnamefont {A.}~\bibnamefont {Remm}}, \bibinfo {author} {\bibfnamefont
  {S.}~\bibnamefont {Lazar}}, \bibinfo {author} {\bibfnamefont {J.-C.}\
  \bibnamefont {Besse}}, \bibinfo {author} {\bibfnamefont {T.}~\bibnamefont
  {Walter}}, \bibinfo {author} {\bibfnamefont {A.}~\bibnamefont {Wallraff}},\
  and\ \bibinfo {author} {\bibfnamefont {C.}~\bibnamefont {Eichler}},\
  }\bibfield  {title} {\bibinfo {title} {Implementation of conditional phase
  gates based on tunable $zz$ interactions},\ }\href
  {https://doi.org/10.1103/PhysRevLett.125.240502} {\bibfield  {journal}
  {\bibinfo  {journal} {Phys. Rev. Lett.}\ }\textbf {\bibinfo {volume} {125}},\
  \bibinfo {pages} {240502} (\bibinfo {year} {2020})}\BibitemShut {NoStop}%
\bibitem [{\citenamefont {Ozfidan}\ \emph {et~al.}(2020)\citenamefont
  {Ozfidan}, \citenamefont {Deng}, \citenamefont {Smirnov}, \citenamefont
  {Lanting}, \citenamefont {Harris}, \citenamefont {Swenson}, \citenamefont
  {Whittaker}, \citenamefont {Altomare}, \citenamefont {Babcock}, \citenamefont
  {Baron}, \citenamefont {Berkley}, \citenamefont {Boothby}, \citenamefont
  {Christiani}, \citenamefont {Bunyk}, \citenamefont {Enderud}, \citenamefont
  {Evert}, \citenamefont {Hager}, \citenamefont {Hajda}, \citenamefont
  {Hilton}, \citenamefont {Huang}, \citenamefont {Hoskinson}, \citenamefont
  {Johnson}, \citenamefont {Jooya}, \citenamefont {Ladizinsky}, \citenamefont
  {Ladizinsky}, \citenamefont {Li}, \citenamefont {MacDonald}, \citenamefont
  {Marsden}, \citenamefont {Marsden}, \citenamefont {Medina}, \citenamefont
  {Molavi}, \citenamefont {Neufeld}, \citenamefont {Nissen}, \citenamefont
  {Norouzpour}, \citenamefont {Oh}, \citenamefont {Pavlov}, \citenamefont
  {Perminov}, \citenamefont {Poulin-Lamarre}, \citenamefont {Reis},
  \citenamefont {Prescott}, \citenamefont {Rich}, \citenamefont {Sato},
  \citenamefont {Sterling}, \citenamefont {Tsai}, \citenamefont {Volkmann},
  \citenamefont {Wilkinson}, \citenamefont {Yao},\ and\ \citenamefont
  {Amin}}]{ozfidan2020}%
  \BibitemOpen
  \bibfield  {author} {\bibinfo {author} {\bibfnamefont {I.}~\bibnamefont
  {Ozfidan}}, \bibinfo {author} {\bibfnamefont {C.}~\bibnamefont {Deng}},
  \bibinfo {author} {\bibfnamefont {A.}~\bibnamefont {Smirnov}}, \bibinfo
  {author} {\bibfnamefont {T.}~\bibnamefont {Lanting}}, \bibinfo {author}
  {\bibfnamefont {R.}~\bibnamefont {Harris}}, \bibinfo {author} {\bibfnamefont
  {L.}~\bibnamefont {Swenson}}, \bibinfo {author} {\bibfnamefont
  {J.}~\bibnamefont {Whittaker}}, \bibinfo {author} {\bibfnamefont
  {F.}~\bibnamefont {Altomare}}, \bibinfo {author} {\bibfnamefont
  {M.}~\bibnamefont {Babcock}}, \bibinfo {author} {\bibfnamefont
  {C.}~\bibnamefont {Baron}}, \bibinfo {author} {\bibfnamefont
  {A.}~\bibnamefont {Berkley}}, \bibinfo {author} {\bibfnamefont
  {K.}~\bibnamefont {Boothby}}, \bibinfo {author} {\bibfnamefont
  {H.}~\bibnamefont {Christiani}}, \bibinfo {author} {\bibfnamefont
  {P.}~\bibnamefont {Bunyk}}, \bibinfo {author} {\bibfnamefont
  {C.}~\bibnamefont {Enderud}}, \bibinfo {author} {\bibfnamefont
  {B.}~\bibnamefont {Evert}}, \bibinfo {author} {\bibfnamefont
  {M.}~\bibnamefont {Hager}}, \bibinfo {author} {\bibfnamefont
  {A.}~\bibnamefont {Hajda}}, \bibinfo {author} {\bibfnamefont
  {J.}~\bibnamefont {Hilton}}, \bibinfo {author} {\bibfnamefont
  {S.}~\bibnamefont {Huang}}, \bibinfo {author} {\bibfnamefont
  {E.}~\bibnamefont {Hoskinson}}, \bibinfo {author} {\bibfnamefont
  {M.}~\bibnamefont {Johnson}}, \bibinfo {author} {\bibfnamefont
  {K.}~\bibnamefont {Jooya}}, \bibinfo {author} {\bibfnamefont
  {E.}~\bibnamefont {Ladizinsky}}, \bibinfo {author} {\bibfnamefont
  {N.}~\bibnamefont {Ladizinsky}}, \bibinfo {author} {\bibfnamefont
  {R.}~\bibnamefont {Li}}, \bibinfo {author} {\bibfnamefont {A.}~\bibnamefont
  {MacDonald}}, \bibinfo {author} {\bibfnamefont {D.}~\bibnamefont {Marsden}},
  \bibinfo {author} {\bibfnamefont {G.}~\bibnamefont {Marsden}}, \bibinfo
  {author} {\bibfnamefont {T.}~\bibnamefont {Medina}}, \bibinfo {author}
  {\bibfnamefont {R.}~\bibnamefont {Molavi}}, \bibinfo {author} {\bibfnamefont
  {R.}~\bibnamefont {Neufeld}}, \bibinfo {author} {\bibfnamefont
  {M.}~\bibnamefont {Nissen}}, \bibinfo {author} {\bibfnamefont
  {M.}~\bibnamefont {Norouzpour}}, \bibinfo {author} {\bibfnamefont
  {T.}~\bibnamefont {Oh}}, \bibinfo {author} {\bibfnamefont {I.}~\bibnamefont
  {Pavlov}}, \bibinfo {author} {\bibfnamefont {I.}~\bibnamefont {Perminov}},
  \bibinfo {author} {\bibfnamefont {G.}~\bibnamefont {Poulin-Lamarre}},
  \bibinfo {author} {\bibfnamefont {M.}~\bibnamefont {Reis}}, \bibinfo {author}
  {\bibfnamefont {T.}~\bibnamefont {Prescott}}, \bibinfo {author}
  {\bibfnamefont {C.}~\bibnamefont {Rich}}, \bibinfo {author} {\bibfnamefont
  {Y.}~\bibnamefont {Sato}}, \bibinfo {author} {\bibfnamefont {G.}~\bibnamefont
  {Sterling}}, \bibinfo {author} {\bibfnamefont {N.}~\bibnamefont {Tsai}},
  \bibinfo {author} {\bibfnamefont {M.}~\bibnamefont {Volkmann}}, \bibinfo
  {author} {\bibfnamefont {W.}~\bibnamefont {Wilkinson}}, \bibinfo {author}
  {\bibfnamefont {J.}~\bibnamefont {Yao}},\ and\ \bibinfo {author}
  {\bibfnamefont {M.}~\bibnamefont {Amin}},\ }\bibfield  {title} {\bibinfo
  {title} {Demonstration of a nonstoquastic hamiltonian in coupled
  superconducting flux qubits},\ }\href
  {https://doi.org/10.1103/PhysRevApplied.13.034037} {\bibfield  {journal}
  {\bibinfo  {journal} {Phys. Rev. Applied}\ }\textbf {\bibinfo {volume}
  {13}},\ \bibinfo {pages} {034037} (\bibinfo {year} {2020})}\BibitemShut
  {NoStop}%
\bibitem [{\citenamefont {Consani}\ and\ \citenamefont
  {Warburton}(2020)}]{Consani2020}%
  \BibitemOpen
  \bibfield  {author} {\bibinfo {author} {\bibfnamefont {G.}~\bibnamefont
  {Consani}}\ and\ \bibinfo {author} {\bibfnamefont {P.~A.}\ \bibnamefont
  {Warburton}},\ }\bibfield  {title} {\bibinfo {title} {Effective hamiltonians
  for interacting superconducting qubits: local basis reduction and the
  schrieffer–wolff transformation},\ }\href
  {https://doi.org/10.1088/1367-2630/ab83d1} {\bibfield  {journal} {\bibinfo
  {journal} {New Journal of Physics}\ }\textbf {\bibinfo {volume} {22}},\
  \bibinfo {pages} {053040} (\bibinfo {year} {2020})}\BibitemShut {NoStop}%
\bibitem [{\citenamefont {Kerman}(2019)}]{Kerman2019}%
  \BibitemOpen
  \bibfield  {author} {\bibinfo {author} {\bibfnamefont {A.~J.}\ \bibnamefont
  {Kerman}},\ }\bibfield  {title} {\bibinfo {title} {Superconducting qubit
  circuit emulation of a vector spin-1/2},\ }\href
  {https://doi.org/10.1088/1367-2630/ab2ee7} {\bibfield  {journal} {\bibinfo
  {journal} {New Journal of Physics}\ }\textbf {\bibinfo {volume} {21}},\
  \bibinfo {pages} {073030} (\bibinfo {year} {2019})}\BibitemShut {NoStop}%
\bibitem [{\citenamefont {Arute}\ \emph {et~al.}(2019)\citenamefont {Arute},
  \citenamefont {Arya}, \citenamefont {Babbush}, \citenamefont {Bacon},
  \citenamefont {Bardin}, \citenamefont {Barends}, \citenamefont {Biswas},
  \citenamefont {Boixo}, \citenamefont {Brandao}, \citenamefont {Buell},
  \citenamefont {Burkett}, \citenamefont {Chen}, \citenamefont {Chen},
  \citenamefont {Chiaro}, \citenamefont {Collins}, \citenamefont {Courtney},
  \citenamefont {Dunsworth}, \citenamefont {Farhi}, \citenamefont {Foxen},
  \citenamefont {Fowler}, \citenamefont {Gidney}, \citenamefont {Giustina},
  \citenamefont {Graff}, \citenamefont {Guerin}, \citenamefont {Habegger},
  \citenamefont {Harrigan}, \citenamefont {Hartmann}, \citenamefont {Ho},
  \citenamefont {Hoffmann}, \citenamefont {Huang}, \citenamefont {Humble},
  \citenamefont {Isakov}, \citenamefont {Jeffrey}, \citenamefont {Jiang},
  \citenamefont {Kafri}, \citenamefont {Kechedzhi}, \citenamefont {Kelly},
  \citenamefont {Klimov}, \citenamefont {Knysh}, \citenamefont {Korotkov},
  \citenamefont {Kostritsa}, \citenamefont {Landhuis}, \citenamefont
  {Lindmark}, \citenamefont {Lucero}, \citenamefont {Lyakh}, \citenamefont
  {Mandrà}, \citenamefont {McClean}, \citenamefont {McEwen}, \citenamefont
  {Megrant}, \citenamefont {Mi}, \citenamefont {Michielsen}, \citenamefont
  {Mohseni}, \citenamefont {Mutus}, \citenamefont {Naaman}, \citenamefont
  {Neeley}, \citenamefont {Neill}, \citenamefont {Yuezhen~Niu}, \citenamefont
  {Ostby}, \citenamefont {Petukhov}, \citenamefont {Platt}, \citenamefont
  {Quintana}, \citenamefont {Rieffel}, \citenamefont {Roushan}, \citenamefont
  {Rubin}, \citenamefont {Sank}, \citenamefont {Satzinger}, \citenamefont
  {Smelyanskiy}, \citenamefont {Sung}, \citenamefont {Trevithick},
  \citenamefont {Vainsencher}, \citenamefont {Villalonga}, \citenamefont
  {White}, \citenamefont {Yao}, \citenamefont {Yeh}, \citenamefont {Zalcman},\
  and\ \citenamefont {Neven}}]{arute2019quantum}%
  \BibitemOpen
  \bibfield  {author} {\bibinfo {author} {\bibfnamefont {F.}~\bibnamefont
  {Arute}}, \bibinfo {author} {\bibfnamefont {K.}~\bibnamefont {Arya}},
  \bibinfo {author} {\bibfnamefont {R.}~\bibnamefont {Babbush}}, \bibinfo
  {author} {\bibfnamefont {D.}~\bibnamefont {Bacon}}, \bibinfo {author}
  {\bibfnamefont {J.~C.}\ \bibnamefont {Bardin}}, \bibinfo {author}
  {\bibfnamefont {R.}~\bibnamefont {Barends}}, \bibinfo {author} {\bibfnamefont
  {R.}~\bibnamefont {Biswas}}, \bibinfo {author} {\bibfnamefont
  {S.}~\bibnamefont {Boixo}}, \bibinfo {author} {\bibfnamefont {F.~G. S.~L.}\
  \bibnamefont {Brandao}}, \bibinfo {author} {\bibfnamefont {D.~A.}\
  \bibnamefont {Buell}}, \bibinfo {author} {\bibfnamefont {B.}~\bibnamefont
  {Burkett}}, \bibinfo {author} {\bibfnamefont {Y.}~\bibnamefont {Chen}},
  \bibinfo {author} {\bibfnamefont {Z.}~\bibnamefont {Chen}}, \bibinfo {author}
  {\bibfnamefont {B.}~\bibnamefont {Chiaro}}, \bibinfo {author} {\bibfnamefont
  {R.}~\bibnamefont {Collins}}, \bibinfo {author} {\bibfnamefont
  {W.}~\bibnamefont {Courtney}}, \bibinfo {author} {\bibfnamefont
  {A.}~\bibnamefont {Dunsworth}}, \bibinfo {author} {\bibfnamefont
  {E.}~\bibnamefont {Farhi}}, \bibinfo {author} {\bibfnamefont
  {B.}~\bibnamefont {Foxen}}, \bibinfo {author} {\bibfnamefont
  {A.}~\bibnamefont {Fowler}}, \bibinfo {author} {\bibfnamefont
  {C.}~\bibnamefont {Gidney}}, \bibinfo {author} {\bibfnamefont
  {M.}~\bibnamefont {Giustina}}, \bibinfo {author} {\bibfnamefont
  {R.}~\bibnamefont {Graff}}, \bibinfo {author} {\bibfnamefont
  {K.}~\bibnamefont {Guerin}}, \bibinfo {author} {\bibfnamefont
  {S.}~\bibnamefont {Habegger}}, \bibinfo {author} {\bibfnamefont
  {M.}~\bibnamefont {Harrigan}}, \bibinfo {author} {\bibfnamefont {M.~J.}\
  \bibnamefont {Hartmann}}, \bibinfo {author} {\bibfnamefont {A.}~\bibnamefont
  {Ho}}, \bibinfo {author} {\bibfnamefont {M.}~\bibnamefont {Hoffmann}},
  \bibinfo {author} {\bibfnamefont {T.}~\bibnamefont {Huang}}, \bibinfo
  {author} {\bibfnamefont {T.~S.}\ \bibnamefont {Humble}}, \bibinfo {author}
  {\bibfnamefont {S.~V.}\ \bibnamefont {Isakov}}, \bibinfo {author}
  {\bibfnamefont {E.}~\bibnamefont {Jeffrey}}, \bibinfo {author} {\bibfnamefont
  {Z.}~\bibnamefont {Jiang}}, \bibinfo {author} {\bibfnamefont
  {D.}~\bibnamefont {Kafri}}, \bibinfo {author} {\bibfnamefont
  {K.}~\bibnamefont {Kechedzhi}}, \bibinfo {author} {\bibfnamefont
  {J.}~\bibnamefont {Kelly}}, \bibinfo {author} {\bibfnamefont {P.~V.}\
  \bibnamefont {Klimov}}, \bibinfo {author} {\bibfnamefont {S.}~\bibnamefont
  {Knysh}}, \bibinfo {author} {\bibfnamefont {A.}~\bibnamefont {Korotkov}},
  \bibinfo {author} {\bibfnamefont {F.}~\bibnamefont {Kostritsa}}, \bibinfo
  {author} {\bibfnamefont {D.}~\bibnamefont {Landhuis}}, \bibinfo {author}
  {\bibfnamefont {M.}~\bibnamefont {Lindmark}}, \bibinfo {author}
  {\bibfnamefont {E.}~\bibnamefont {Lucero}}, \bibinfo {author} {\bibfnamefont
  {D.}~\bibnamefont {Lyakh}}, \bibinfo {author} {\bibfnamefont
  {S.}~\bibnamefont {Mandrà}}, \bibinfo {author} {\bibfnamefont {J.~R.}\
  \bibnamefont {McClean}}, \bibinfo {author} {\bibfnamefont {M.}~\bibnamefont
  {McEwen}}, \bibinfo {author} {\bibfnamefont {A.}~\bibnamefont {Megrant}},
  \bibinfo {author} {\bibfnamefont {X.}~\bibnamefont {Mi}}, \bibinfo {author}
  {\bibfnamefont {K.}~\bibnamefont {Michielsen}}, \bibinfo {author}
  {\bibfnamefont {M.}~\bibnamefont {Mohseni}}, \bibinfo {author} {\bibfnamefont
  {J.}~\bibnamefont {Mutus}}, \bibinfo {author} {\bibfnamefont
  {O.}~\bibnamefont {Naaman}}, \bibinfo {author} {\bibfnamefont
  {M.}~\bibnamefont {Neeley}}, \bibinfo {author} {\bibfnamefont
  {C.}~\bibnamefont {Neill}}, \bibinfo {author} {\bibfnamefont
  {M.}~\bibnamefont {Yuezhen~Niu}}, \bibinfo {author} {\bibfnamefont
  {E.}~\bibnamefont {Ostby}}, \bibinfo {author} {\bibfnamefont
  {A.}~\bibnamefont {Petukhov}}, \bibinfo {author} {\bibfnamefont {J.~C.}\
  \bibnamefont {Platt}}, \bibinfo {author} {\bibfnamefont {C.}~\bibnamefont
  {Quintana}}, \bibinfo {author} {\bibfnamefont {E.~G.}\ \bibnamefont
  {Rieffel}}, \bibinfo {author} {\bibfnamefont {P.}~\bibnamefont {Roushan}},
  \bibinfo {author} {\bibfnamefont {N.~C.}\ \bibnamefont {Rubin}}, \bibinfo
  {author} {\bibfnamefont {D.}~\bibnamefont {Sank}}, \bibinfo {author}
  {\bibfnamefont {K.~J.}\ \bibnamefont {Satzinger}}, \bibinfo {author}
  {\bibfnamefont {V.}~\bibnamefont {Smelyanskiy}}, \bibinfo {author}
  {\bibfnamefont {K.~J.}\ \bibnamefont {Sung}}, \bibinfo {author}
  {\bibfnamefont {M.~D.}\ \bibnamefont {Trevithick}}, \bibinfo {author}
  {\bibfnamefont {A.}~\bibnamefont {Vainsencher}}, \bibinfo {author}
  {\bibfnamefont {B.}~\bibnamefont {Villalonga}}, \bibinfo {author}
  {\bibfnamefont {T.}~\bibnamefont {White}}, \bibinfo {author} {\bibfnamefont
  {Z.~J.}\ \bibnamefont {Yao}}, \bibinfo {author} {\bibfnamefont
  {P.}~\bibnamefont {Yeh}}, \bibinfo {author} {\bibfnamefont {A.}~\bibnamefont
  {Zalcman}},\ and\ \bibinfo {author} {\bibfnamefont {H.}~\bibnamefont
  {Neven}},\ }\bibfield  {title} {\bibinfo {title} {Quantum supremacy using a
  programmable superconducting processor},\ }\href
  {https://doi.org/10.1038/s41586-019-1666-5} {\bibfield  {journal} {\bibinfo
  {journal} {Nature}\ }\textbf {\bibinfo {volume} {574}},\ \bibinfo {pages}
  {505} (\bibinfo {year} {2019})}\BibitemShut {NoStop}%
\bibitem [{\citenamefont {Bravyi}\ \emph {et~al.}(2008)\citenamefont {Bravyi},
  \citenamefont {Divincenzo}, \citenamefont {Oliveira},\ and\ \citenamefont
  {Terhal}}]{bravyi2006}%
  \BibitemOpen
  \bibfield  {author} {\bibinfo {author} {\bibfnamefont {S.}~\bibnamefont
  {Bravyi}}, \bibinfo {author} {\bibfnamefont {D.~P.}\ \bibnamefont
  {Divincenzo}}, \bibinfo {author} {\bibfnamefont {R.}~\bibnamefont
  {Oliveira}},\ and\ \bibinfo {author} {\bibfnamefont {B.~M.}\ \bibnamefont
  {Terhal}},\ }\bibfield  {title} {\bibinfo {title} {The complexity of
  stoquastic local hamiltonian problems},\ }\href@noop {} {\bibfield  {journal}
  {\bibinfo  {journal} {Quantum Info. Comput.}\ }\textbf {\bibinfo {volume}
  {8}},\ \bibinfo {pages} {361} (\bibinfo {year} {2008})}\BibitemShut {NoStop}%
\bibitem [{\citenamefont {Susa}\ \emph {et~al.}(2017)\citenamefont {Susa},
  \citenamefont {Jadebeck},\ and\ \citenamefont {Nishimori}}]{susa2017}%
  \BibitemOpen
  \bibfield  {author} {\bibinfo {author} {\bibfnamefont {Y.}~\bibnamefont
  {Susa}}, \bibinfo {author} {\bibfnamefont {J.~F.}\ \bibnamefont {Jadebeck}},\
  and\ \bibinfo {author} {\bibfnamefont {H.}~\bibnamefont {Nishimori}},\
  }\bibfield  {title} {\bibinfo {title} {Relation between quantum fluctuations
  and the performance enhancement of quantum annealing in a nonstoquastic
  hamiltonian},\ }\href {https://doi.org/10.1103/physreva.95.042321} {\bibfield
   {journal} {\bibinfo  {journal} {Physical Review A}\ }\textbf {\bibinfo
  {volume} {95}},\ \bibinfo {pages} {042321} (\bibinfo {year}
  {2017})}\BibitemShut {NoStop}%
\bibitem [{\citenamefont {Albash}(2019)}]{albash2019role}%
  \BibitemOpen
  \bibfield  {author} {\bibinfo {author} {\bibfnamefont {T.}~\bibnamefont
  {Albash}},\ }\bibfield  {title} {\bibinfo {title} {Role of nonstoquastic
  catalysts in quantum adiabatic optimization},\ }\href
  {https://doi.org/10.1103/physreva.99.042334} {\bibfield  {journal} {\bibinfo
  {journal} {Physical Review A}\ }\textbf {\bibinfo {volume} {99}},\ \bibinfo
  {pages} {042334} (\bibinfo {year} {2019})}\BibitemShut {NoStop}%
\bibitem [{\citenamefont {Hormozi}\ \emph {et~al.}(2017)\citenamefont
  {Hormozi}, \citenamefont {Brown}, \citenamefont {Carleo},\ and\ \citenamefont
  {Troyer}}]{hormozi2017}%
  \BibitemOpen
  \bibfield  {author} {\bibinfo {author} {\bibfnamefont {L.}~\bibnamefont
  {Hormozi}}, \bibinfo {author} {\bibfnamefont {E.~W.}\ \bibnamefont {Brown}},
  \bibinfo {author} {\bibfnamefont {G.}~\bibnamefont {Carleo}},\ and\ \bibinfo
  {author} {\bibfnamefont {M.}~\bibnamefont {Troyer}},\ }\bibfield  {title}
  {\bibinfo {title} {Nonstoquastic hamiltonians and quantum annealing of an
  ising spin glass},\ }\href {https://doi.org/10.1103/physrevb.95.184416}
  {\bibfield  {journal} {\bibinfo  {journal} {Physical review B}\ }\textbf
  {\bibinfo {volume} {95}},\ \bibinfo {pages} {184416} (\bibinfo {year}
  {2017})}\BibitemShut {NoStop}%
\bibitem [{\citenamefont {Albash}\ and\ \citenamefont
  {Lidar}(2018)}]{albash2018adiabatic}%
  \BibitemOpen
  \bibfield  {author} {\bibinfo {author} {\bibfnamefont {T.}~\bibnamefont
  {Albash}}\ and\ \bibinfo {author} {\bibfnamefont {D.~A.}\ \bibnamefont
  {Lidar}},\ }\bibfield  {title} {\bibinfo {title} {Adiabatic quantum
  computation},\ }\href@noop {} {\bibfield  {journal} {\bibinfo  {journal}
  {Reviews of Modern Physics}\ }\textbf {\bibinfo {volume} {90}},\ \bibinfo
  {pages} {015002} (\bibinfo {year} {2018})}\BibitemShut {NoStop}%
\bibitem [{\citenamefont {Kempe}\ \emph {et~al.}(2006)\citenamefont {Kempe},
  \citenamefont {Kitaev},\ and\ \citenamefont {Regev}}]{kempe2006complexity}%
  \BibitemOpen
  \bibfield  {author} {\bibinfo {author} {\bibfnamefont {J.}~\bibnamefont
  {Kempe}}, \bibinfo {author} {\bibfnamefont {A.}~\bibnamefont {Kitaev}},\ and\
  \bibinfo {author} {\bibfnamefont {O.}~\bibnamefont {Regev}},\ }\bibfield
  {title} {\bibinfo {title} {The complexity of the local hamiltonian problem},\
  }\href {https://doi.org/10.1137/S0097539704445226} {\bibfield  {journal}
  {\bibinfo  {journal} {SIAM Journal on Computing}\ }\textbf {\bibinfo {volume}
  {35}},\ \bibinfo {pages} {1070} (\bibinfo {year} {2006})}\BibitemShut
  {NoStop}%
\bibitem [{\citenamefont {Oliveira}\ and\ \citenamefont
  {Terhal}(2008)}]{oliveira2005complexity}%
  \BibitemOpen
  \bibfield  {author} {\bibinfo {author} {\bibfnamefont {R.}~\bibnamefont
  {Oliveira}}\ and\ \bibinfo {author} {\bibfnamefont {B.~M.}\ \bibnamefont
  {Terhal}},\ }\bibfield  {title} {\bibinfo {title} {The complexity of quantum
  spin systems on a two-dimensional square lattice},\ }\href@noop {} {\bibfield
   {journal} {\bibinfo  {journal} {Quantum Info. Comput.}\ }\textbf {\bibinfo
  {volume} {8}},\ \bibinfo {pages} {900–924} (\bibinfo {year}
  {2008})}\BibitemShut {NoStop}%
\bibitem [{\citenamefont {Orlando}\ \emph {et~al.}(1999)\citenamefont
  {Orlando}, \citenamefont {Mooij}, \citenamefont {Tian}, \citenamefont
  {van~der Wal}, \citenamefont {Levitov}, \citenamefont {Lloyd},\ and\
  \citenamefont {Mazo}}]{orlando}%
  \BibitemOpen
  \bibfield  {author} {\bibinfo {author} {\bibfnamefont {T.~P.}\ \bibnamefont
  {Orlando}}, \bibinfo {author} {\bibfnamefont {J.~E.}\ \bibnamefont {Mooij}},
  \bibinfo {author} {\bibfnamefont {L.}~\bibnamefont {Tian}}, \bibinfo {author}
  {\bibfnamefont {C.~H.}\ \bibnamefont {van~der Wal}}, \bibinfo {author}
  {\bibfnamefont {L.~S.}\ \bibnamefont {Levitov}}, \bibinfo {author}
  {\bibfnamefont {S.}~\bibnamefont {Lloyd}},\ and\ \bibinfo {author}
  {\bibfnamefont {J.~J.}\ \bibnamefont {Mazo}},\ }\bibfield  {title} {\bibinfo
  {title} {Superconducting persistent-current qubit},\ }\href
  {https://doi.org/10.1103/PhysRevB.60.15398} {\bibfield  {journal} {\bibinfo
  {journal} {Phys. Rev. B}\ }\textbf {\bibinfo {volume} {60}},\ \bibinfo
  {pages} {15398} (\bibinfo {year} {1999})}\BibitemShut {NoStop}%
\bibitem [{\citenamefont {Satoh}\ \emph {et~al.}(2015)\citenamefont {Satoh},
  \citenamefont {Matsuzaki}, \citenamefont {Kakuyanagi}, \citenamefont {Semba},
  \citenamefont {Yamaguchi},\ and\ \citenamefont {Saito}}]{satoh2015ising}%
  \BibitemOpen
  \bibfield  {author} {\bibinfo {author} {\bibfnamefont {T.}~\bibnamefont
  {Satoh}}, \bibinfo {author} {\bibfnamefont {Y.}~\bibnamefont {Matsuzaki}},
  \bibinfo {author} {\bibfnamefont {K.}~\bibnamefont {Kakuyanagi}}, \bibinfo
  {author} {\bibfnamefont {K.}~\bibnamefont {Semba}}, \bibinfo {author}
  {\bibfnamefont {H.}~\bibnamefont {Yamaguchi}},\ and\ \bibinfo {author}
  {\bibfnamefont {S.}~\bibnamefont {Saito}},\ }\href@noop {} {\bibinfo {title}
  {Ising interaction between capacitively-coupled superconducting flux qubits}}
  (\bibinfo {year} {2015}),\ \Eprint {https://arxiv.org/abs/1501.07739}
  {arXiv:1501.07739 [quant-ph]} \BibitemShut {NoStop}%
\bibitem [{\citenamefont {Hita-Pérez}\ \emph {et~al.}(2021)\citenamefont
  {Hita-Pérez}, \citenamefont {Jaumà}, \citenamefont {Pino},\ and\
  \citenamefont {García-Ripoll}}]{hitaperez2021ultrastrong}%
  \BibitemOpen
  \bibfield  {author} {\bibinfo {author} {\bibfnamefont {M.}~\bibnamefont
  {Hita-Pérez}}, \bibinfo {author} {\bibfnamefont {G.}~\bibnamefont {Jaumà}},
  \bibinfo {author} {\bibfnamefont {M.}~\bibnamefont {Pino}},\ and\ \bibinfo
  {author} {\bibfnamefont {J.~J.}\ \bibnamefont {García-Ripoll}},\ }\href@noop
  {} {\bibinfo {title} {Ultrastrong capacitive coupling of flux qubits}}
  (\bibinfo {year} {2021}),\ \Eprint {https://arxiv.org/abs/2108.02549}
  {arXiv:2108.02549 [quant-ph]} \BibitemShut {NoStop}%
\bibitem [{\citenamefont {Chiorescu}\ \emph {et~al.}(2003)\citenamefont
  {Chiorescu}, \citenamefont {Nakamura}, \citenamefont {Harmans},\ and\
  \citenamefont {Mooij}}]{Chiorescu_2003}%
  \BibitemOpen
  \bibfield  {author} {\bibinfo {author} {\bibfnamefont {I.}~\bibnamefont
  {Chiorescu}}, \bibinfo {author} {\bibfnamefont {Y.}~\bibnamefont {Nakamura}},
  \bibinfo {author} {\bibfnamefont {C.~J. P.~M.}\ \bibnamefont {Harmans}},\
  and\ \bibinfo {author} {\bibfnamefont {J.~E.}\ \bibnamefont {Mooij}},\
  }\bibfield  {title} {\bibinfo {title} {Coherent quantum dynamics of a
  superconducting flux qubit},\ }\href
  {https://doi.org/10.1126/science.1081045} {\bibfield  {journal} {\bibinfo
  {journal} {Science}\ }\textbf {\bibinfo {volume} {299}},\ \bibinfo {pages}
  {1869} (\bibinfo {year} {2003})}\BibitemShut {NoStop}%
\bibitem [{\citenamefont {Schrieffer}\ and\ \citenamefont
  {Wolff}(1966)}]{schrieffer1966relation}%
  \BibitemOpen
  \bibfield  {author} {\bibinfo {author} {\bibfnamefont {J.~R.}\ \bibnamefont
  {Schrieffer}}\ and\ \bibinfo {author} {\bibfnamefont {P.~A.}\ \bibnamefont
  {Wolff}},\ }\bibfield  {title} {\bibinfo {title} {Relation between the
  anderson and kondo hamiltonians},\ }\href
  {https://doi.org/10.1103/PhysRev.149.491} {\bibfield  {journal} {\bibinfo
  {journal} {Physical Review}\ }\textbf {\bibinfo {volume} {149}},\ \bibinfo
  {pages} {491} (\bibinfo {year} {1966})}\BibitemShut {NoStop}%
\bibitem [{\citenamefont {Bravyi}\ \emph {et~al.}(2011)\citenamefont {Bravyi},
  \citenamefont {DiVincenzo},\ and\ \citenamefont {Loss}}]{bravyi2011}%
  \BibitemOpen
  \bibfield  {author} {\bibinfo {author} {\bibfnamefont {S.}~\bibnamefont
  {Bravyi}}, \bibinfo {author} {\bibfnamefont {D.~P.}\ \bibnamefont
  {DiVincenzo}},\ and\ \bibinfo {author} {\bibfnamefont {D.}~\bibnamefont
  {Loss}},\ }\bibfield  {title} {\bibinfo {title} {Schrieffer--wolff
  transformation for quantum many-body systems},\ }\href
  {https://doi.org/10.1016/j.aop.2011.06.004} {\bibfield  {journal} {\bibinfo
  {journal} {Annals of physics}\ }\textbf {\bibinfo {volume} {326}},\ \bibinfo
  {pages} {2793} (\bibinfo {year} {2011})}\BibitemShut {NoStop}%
\bibitem [{\citenamefont {Golub}\ and\ \citenamefont
  {Van~Loan}(1996)}]{golub1996matrix}%
  \BibitemOpen
  \bibfield  {author} {\bibinfo {author} {\bibfnamefont {G.~H.}\ \bibnamefont
  {Golub}}\ and\ \bibinfo {author} {\bibfnamefont {C.~F.}\ \bibnamefont
  {Van~Loan}},\ }\href@noop {} {\emph {\bibinfo {title} {Matrix Computations
  (3rd Ed.)}}}\ (\bibinfo  {publisher} {Johns Hopkins University Press},\
  \bibinfo {address} {Baltimore},\ \bibinfo {year} {1996})\BibitemShut
  {NoStop}%
\bibitem [{\citenamefont {van~der Wal}\ \emph {et~al.}(2000)\citenamefont
  {van~der Wal}, \citenamefont {ter Haar}, \citenamefont {Wilhelm},
  \citenamefont {Schouten}, \citenamefont {Harmans}, \citenamefont {Orlando},
  \citenamefont {Lloyd},\ and\ \citenamefont {Mooij}}]{orlando2}%
  \BibitemOpen
  \bibfield  {author} {\bibinfo {author} {\bibfnamefont {C.~H.}\ \bibnamefont
  {van~der Wal}}, \bibinfo {author} {\bibfnamefont {A.~C.~J.}\ \bibnamefont
  {ter Haar}}, \bibinfo {author} {\bibfnamefont {F.~K.}\ \bibnamefont
  {Wilhelm}}, \bibinfo {author} {\bibfnamefont {R.~N.}\ \bibnamefont
  {Schouten}}, \bibinfo {author} {\bibfnamefont {C.~J. P.~M.}\ \bibnamefont
  {Harmans}}, \bibinfo {author} {\bibfnamefont {T.~P.}\ \bibnamefont
  {Orlando}}, \bibinfo {author} {\bibfnamefont {S.}~\bibnamefont {Lloyd}},\
  and\ \bibinfo {author} {\bibfnamefont {J.~E.}\ \bibnamefont {Mooij}},\
  }\bibfield  {title} {\bibinfo {title} {Quantum superposition of macroscopic
  persistent-current states},\ }\href
  {https://doi.org/10.1126/science.290.5492.773} {\bibfield  {journal}
  {\bibinfo  {journal} {Science}\ }\textbf {\bibinfo {volume} {290}},\ \bibinfo
  {pages} {773} (\bibinfo {year} {2000})}\BibitemShut {NoStop}%
\bibitem [{\citenamefont {Yan}\ \emph {et~al.}(2016)\citenamefont {Yan},
  \citenamefont {Gustavsson}, \citenamefont {Kamal}, \citenamefont {Birenbaum},
  \citenamefont {Sears}, \citenamefont {Hover}, \citenamefont {Gudmundsen},
  \citenamefont {Rosenberg}, \citenamefont {Samach}, \citenamefont {Weber}
  \emph {et~al.}}]{yan2016}%
  \BibitemOpen
  \bibfield  {author} {\bibinfo {author} {\bibfnamefont {F.}~\bibnamefont
  {Yan}}, \bibinfo {author} {\bibfnamefont {S.}~\bibnamefont {Gustavsson}},
  \bibinfo {author} {\bibfnamefont {A.}~\bibnamefont {Kamal}}, \bibinfo
  {author} {\bibfnamefont {J.}~\bibnamefont {Birenbaum}}, \bibinfo {author}
  {\bibfnamefont {A.~P.}\ \bibnamefont {Sears}}, \bibinfo {author}
  {\bibfnamefont {D.}~\bibnamefont {Hover}}, \bibinfo {author} {\bibfnamefont
  {T.~J.}\ \bibnamefont {Gudmundsen}}, \bibinfo {author} {\bibfnamefont
  {D.}~\bibnamefont {Rosenberg}}, \bibinfo {author} {\bibfnamefont
  {G.}~\bibnamefont {Samach}}, \bibinfo {author} {\bibfnamefont
  {S.}~\bibnamefont {Weber}}, \emph {et~al.},\ }\bibfield  {title} {\bibinfo
  {title} {The flux qubit revisited to enhance coherence and reproducibility},\
  }\href@noop {} {\bibfield  {journal} {\bibinfo  {journal} {Nature
  communications}\ }\textbf {\bibinfo {volume} {7}},\ \bibinfo {pages} {1}
  (\bibinfo {year} {2016})}\BibitemShut {NoStop}%
\bibitem [{\citenamefont {Ciani}\ and\ \citenamefont
  {Terhal}(2021)}]{ciani2020}%
  \BibitemOpen
  \bibfield  {author} {\bibinfo {author} {\bibfnamefont {A.}~\bibnamefont
  {Ciani}}\ and\ \bibinfo {author} {\bibfnamefont {B.~M.}\ \bibnamefont
  {Terhal}},\ }\bibfield  {title} {\bibinfo {title} {Stoquasticity in circuit
  qed},\ }\href {https://doi.org/10.1103/PhysRevA.103.042401} {\bibfield
  {journal} {\bibinfo  {journal} {Phys. Rev. A}\ }\textbf {\bibinfo {volume}
  {103}},\ \bibinfo {pages} {042401} (\bibinfo {year} {2021})}\BibitemShut
  {NoStop}%
\bibitem [{\citenamefont {Halverson}\ \emph {et~al.}(2020)\citenamefont
  {Halverson}, \citenamefont {Gupta}, \citenamefont {Goldstein},\ and\
  \citenamefont {Hen}}]{halverson2020efficient}%
  \BibitemOpen
  \bibfield  {author} {\bibinfo {author} {\bibfnamefont {T.}~\bibnamefont
  {Halverson}}, \bibinfo {author} {\bibfnamefont {L.}~\bibnamefont {Gupta}},
  \bibinfo {author} {\bibfnamefont {M.}~\bibnamefont {Goldstein}},\ and\
  \bibinfo {author} {\bibfnamefont {I.}~\bibnamefont {Hen}},\ }\href@noop {}
  {\bibinfo {title} {Efficient simulation of so-called non-stoquastic
  superconducting flux circuits}} (\bibinfo {year} {2020}),\ \Eprint
  {https://arxiv.org/abs/2011.03831} {arXiv:2011.03831 [quant-ph]} \BibitemShut
  {NoStop}%
\bibitem [{\citenamefont {Kafri}\ \emph {et~al.}(2017)\citenamefont {Kafri},
  \citenamefont {Quintana}, \citenamefont {Chen}, \citenamefont {Shabani},
  \citenamefont {Martinis},\ and\ \citenamefont {Neven}}]{kafri}%
  \BibitemOpen
  \bibfield  {author} {\bibinfo {author} {\bibfnamefont {D.}~\bibnamefont
  {Kafri}}, \bibinfo {author} {\bibfnamefont {C.}~\bibnamefont {Quintana}},
  \bibinfo {author} {\bibfnamefont {Y.}~\bibnamefont {Chen}}, \bibinfo {author}
  {\bibfnamefont {A.}~\bibnamefont {Shabani}}, \bibinfo {author} {\bibfnamefont
  {J.~M.}\ \bibnamefont {Martinis}},\ and\ \bibinfo {author} {\bibfnamefont
  {H.}~\bibnamefont {Neven}},\ }\bibfield  {title} {\bibinfo {title} {Tunable
  inductive coupling of superconducting qubits in the strongly nonlinear
  regime},\ }\href {https://doi.org/10.1103/PhysRevA.95.052333} {\bibfield
  {journal} {\bibinfo  {journal} {Phys. Rev. A}\ }\textbf {\bibinfo {volume}
  {95}},\ \bibinfo {pages} {052333} (\bibinfo {year} {2017})}\BibitemShut
  {NoStop}%
\bibitem [{\citenamefont {Weber}\ \emph {et~al.}(2017)\citenamefont {Weber},
  \citenamefont {Samach}, \citenamefont {Hover}, \citenamefont {Gustavsson},
  \citenamefont {Kim}, \citenamefont {Melville}, \citenamefont {Rosenberg},
  \citenamefont {Sears}, \citenamefont {Yan}, \citenamefont {Yoder},\ and\
  \citenamefont {et~al.}}]{Weber_2017}%
  \BibitemOpen
  \bibfield  {author} {\bibinfo {author} {\bibfnamefont {S.~J.}\ \bibnamefont
  {Weber}}, \bibinfo {author} {\bibfnamefont {G.~O.}\ \bibnamefont {Samach}},
  \bibinfo {author} {\bibfnamefont {D.}~\bibnamefont {Hover}}, \bibinfo
  {author} {\bibfnamefont {S.}~\bibnamefont {Gustavsson}}, \bibinfo {author}
  {\bibfnamefont {D.~K.}\ \bibnamefont {Kim}}, \bibinfo {author} {\bibfnamefont
  {A.}~\bibnamefont {Melville}}, \bibinfo {author} {\bibfnamefont
  {D.}~\bibnamefont {Rosenberg}}, \bibinfo {author} {\bibfnamefont {A.~P.}\
  \bibnamefont {Sears}}, \bibinfo {author} {\bibfnamefont {F.}~\bibnamefont
  {Yan}}, \bibinfo {author} {\bibfnamefont {J.~L.}\ \bibnamefont {Yoder}},\
  and\ \bibinfo {author} {\bibnamefont {et~al.}},\ }\bibfield  {title}
  {\bibinfo {title} {Coherent coupled qubits for quantum annealing},\
  }\bibfield  {journal} {\bibinfo  {journal} {Physical Review Applied}\
  }\textbf {\bibinfo {volume} {8}},\ \href
  {https://doi.org/10.1103/physrevapplied.8.014004}
  {10.1103/physrevapplied.8.014004} (\bibinfo {year} {2017})\BibitemShut
  {NoStop}%
\bibitem [{\citenamefont {Allman}\ \emph {et~al.}(2010)\citenamefont {Allman},
  \citenamefont {Altomare}, \citenamefont {Whittaker}, \citenamefont {Cicak},
  \citenamefont {Li}, \citenamefont {Sirois}, \citenamefont {Strong},
  \citenamefont {Teufel},\ and\ \citenamefont {Simmonds}}]{allman2010_a}%
  \BibitemOpen
  \bibfield  {author} {\bibinfo {author} {\bibfnamefont {M.~S.}\ \bibnamefont
  {Allman}}, \bibinfo {author} {\bibfnamefont {F.}~\bibnamefont {Altomare}},
  \bibinfo {author} {\bibfnamefont {J.~D.}\ \bibnamefont {Whittaker}}, \bibinfo
  {author} {\bibfnamefont {K.}~\bibnamefont {Cicak}}, \bibinfo {author}
  {\bibfnamefont {D.}~\bibnamefont {Li}}, \bibinfo {author} {\bibfnamefont
  {A.}~\bibnamefont {Sirois}}, \bibinfo {author} {\bibfnamefont
  {J.}~\bibnamefont {Strong}}, \bibinfo {author} {\bibfnamefont {J.~D.}\
  \bibnamefont {Teufel}},\ and\ \bibinfo {author} {\bibfnamefont {R.~W.}\
  \bibnamefont {Simmonds}},\ }\bibfield  {title} {\bibinfo {title}
  {rf-squid-mediated coherent tunable coupling between a superconducting phase
  qubit and a lumped-element resonator},\ }\href@noop {} {\bibfield  {journal}
  {\bibinfo  {journal} {Physical review letters}\ }\textbf {\bibinfo {volume}
  {104}},\ \bibinfo {pages} {177004} (\bibinfo {year} {2010})}\BibitemShut
  {NoStop}%
\bibitem [{\citenamefont {Devoret}(1997)}]{circuitquantization}%
  \BibitemOpen
  \bibfield  {author} {\bibinfo {author} {\bibfnamefont {M.~H.}\ \bibnamefont
  {Devoret}},\ }\bibfield  {title} {\bibinfo {title} {Quantum fluctuations in
  electrical circuits},\ }in\ \href@noop {} {\emph {\bibinfo {booktitle}
  {Quantum Fluctuations: Les Houches Session LXIII}}},\ \bibinfo {editor}
  {edited by\ \bibinfo {editor} {\bibfnamefont {S.}~\bibnamefont {Reynaud}},
  \bibinfo {editor} {\bibfnamefont {E.}~\bibnamefont {Giacobino}},\ and\
  \bibinfo {editor} {\bibfnamefont {J.}~\bibnamefont {Zinn-Justin}}}\ (\bibinfo
   {publisher} {Elsevier},\ \bibinfo {year} {1997})\ pp.\ \bibinfo {pages}
  {351--386}\BibitemShut {NoStop}%
\end{thebibliography}%

\newpage

\onecolumngrid
\appendix

\clearpage
\setcounter{page}{1}

\renewcommand\thefigure{S.\arabic{figure}}
\renewcommand\theequation{S.\arabic{equation}}

\section*{\label{sec:appendixA} Supplementary Material A:  Extraction of the circuit Hamiltonian}

Following the procedure presented in Ref.~\onlinecite{circuitquantization}  for circuit quantization, we find that the general Lagrangian for the circuits contemplated in this paper reads:
\setcounter{equation}{0}

\begin{equation}
    \label{eq:Lagrangian}
\mathcal{L}=\mathcal{L}_0 +\mathcal{L}_{int}
\end{equation}
where $\mathcal{L}_0=\mathcal{L}_1(\boldsymbol{\phi},\dot{\boldsymbol{\phi}})+\mathcal{L}_2(\boldsymbol{\phi}',\dot{\boldsymbol{\phi}'})$ is the sum of the Lagrangians of the single flux qubits (c-shunted or not) \citep{orlando}
\begin{align}
    \mathcal{L}_q(\boldsymbol{\phi},\dot{\boldsymbol{\phi}})
    =&\frac{1}{2}\dot{\boldsymbol{\phi}}\textbf{C}_q\dot{\boldsymbol{\phi}} + E_J\cos{\left(\frac{\phi_1-\phi_0}{\varphi_0}\right)} + \\
    & + E_J\cos{\left(\frac{\phi_2-\phi_0}{\varphi_0}\right)} + \alpha E_J\cos{\left(\frac{\phi_2-\phi_1-\Phi}{\varphi_0}\right)}\notag
\end{align}
with $E_J$ the characteristic Josephson energy of the junctions,  $\Phi$ the external flux, $\phi_i$ the flux variables, and $\textbf{C}_q$ the capacitance matrix of the flux qubit which depends on the characteristic capacitance of the junctions $C$ and the qubits parameter $\alpha$.
And $\mathcal{L}_{int}$ gives the interaction between qubits and depends on the specific coupling. For a Josephson junction connecting two nodes, $i$ and $j'$, of different flux qubits we obtain a contribution of the form:
\begin{equation}
\mathcal{L}_{int}^{JJ}= \gamma_{ij'}^{JJ} E_J \cos{\left(\frac{\phi'_j-\phi_i}{\varphi_0}\right)}
\end{equation}
While a capacitor coupling gives a contribution:
\begin{equation}
\mathcal{L}_{int}^{cap}= \gamma_{ij'}^{cap} \frac{C}{2} (\dot{\phi'_j}-\dot{\phi_i})^2
\label{eq:capintL}
\end{equation}
In both cases, $\gamma_{ij'}$ represents the proportionality constant, between coupling Josephson energy (capacitance)  and the flux qubits reference Josephson energy, $E_J$ (capacitance, $C$). \\

Using canonical variables, $Q_i=\frac{\partial \mathcal{L}}{\partial\dot{\phi}_i}$, and the Legendre transformation, $H(\mathbf{Q}, \boldsymbol{\phi})=\mathbf{Q}\dot{\boldsymbol{\phi}}-\mathcal{L}(\boldsymbol{\phi},\dot{\boldsymbol{\phi}})$, we can conclude from the previous Lagrangian~\eqref{eq:Lagrangian} that the general Hamiltonian for two coupled 3JJQ is:
\begin{equation}
    H=H_0+H_\text{int}
    \label{eq:Hamiltonian}
\end{equation}
Here,  $H_0=H_1(\mathbf{Q},\boldsymbol{\phi})+H_2(\mathbf{Q}',\boldsymbol{\phi}')$ is the sum of the single flux qubits Hamiltonians (c-shunted or not) whose inverse capacitance matrix may be modified by the action of the coupling  (\textit{renormalization} or \textit{capacitive loading}\citep{Consani2020}), $\widetilde{\mathbf{C}}_q^{-1}$,
\begin{equation}
    \begin{aligned}
        H_q(\mathbf{Q},\boldsymbol{\phi})=\frac{1}{2}\mathbf{Q}\widetilde{\mathbf{C}}_q^{-1}\mathbf{Q}- E_J\cos{\left(\frac{\phi_1-\phi_0}{\varphi_0}\right)} \\
        - E_J\cos{\left(\frac{\phi_2-\phi_0}{\varphi_0}\right)} - \alpha E_J\cos{\left(\frac{\phi_2-\phi_1-\Phi}{\varphi_0}\right)}
    \end{aligned}
\end{equation}
Here ${H}_{int}$ describes the interaction between pairs of different flux qubits. For the inductive coupling mediated by a Josephson junction, we find that
\begin{equation}
    H_\text{int}^{JJ}= - \gamma^{JJ}_{ij'} E_J \cos{\left(\frac{\phi'_j-\phi_i}{\varphi_0}\right)}
\end{equation}
where only the two connected nodes, $i$ and $j$, are implicated and no renormalization for the qubits Hamiltonians has to be considered. However, for the electrostatic interaction mediated by a capacitor we get an interaction term of the form
\begin{equation}
     H_\text{int}^{cap}= \mathbf{Q}\mathbf{C}_c^{-1}\mathbf{Q}'
     \label{eq:capint}
\end{equation}
which gives connections between all node charges of different qubits and depends on the inverse mutual capacitance matrix, $\mathbf{C}_c^{-1}$. The single qubits Hamiltonians are renormalized by rescaling the inverse of the capacitance matrix, $\mathbf{C}_q^{-1}(\alpha)\rightarrow\widetilde{\mathbf{C}}_q^{-1}(\alpha, \gamma^{cap}_{ij})$.\\

To fully understand this procedure it is necessary to define the capacitance matrices and inverse capacitance matrices that have been mentioned during the explanation. The full capacitance matrix for our system (including all nodes in both qubits) is defined as follows
\begin{equation}
    \mathbf{C}=
        \begin{pmatrix}
        \widetilde{\mathbf{C}}_q & -\mathbf{C}_c \\
        -\mathbf{C}^{T}_c & \widetilde{\mathbf{C}}_q'
        \end{pmatrix}.
\end{equation}
For example, the full capacitance matrix for the capacitively coupled circuit in Fig.1(a) is
\begin{equation}
    {\mathbf{C}}=C
        \begin{pmatrix}
        1+\alpha+\beta & -(\alpha+\beta)        &0                       &  0 \\
        -(\alpha+\beta) & 1+\alpha+\beta+\gamma & -\gamma               &  0 \\
        0                & -\gamma             & 1+\alpha+\beta+\gamma  &  -(\alpha+\beta) \\
        0                & 0                     & -(\alpha+\beta)       & 1+\alpha+\beta
        \end{pmatrix}.
\end{equation}
Thus, $\widetilde{\mathbf{C}}_q$ is the renormalized (or not) 3JJQs capacitance matrix whose elements have the form
\begin{equation}
    \begin{aligned}
        & (\widetilde{\mathbf{C}}_q)_{ii}=(\mathbf{C}_q)_{ii}+\sum_{j'}\gamma_{ij'}C \\
        & (\widetilde{\mathbf{C}}_q)_{ij}=(\mathbf{C}_q)_{ij} \\
        & (\widetilde{\mathbf{C}}_q')_{i'i'}=(\mathbf{C}_q)_{i'i'}+\sum_{j}\gamma_{ji'}C \\
        & (\widetilde{\mathbf{C}}_q')_{i'j'}=(\mathbf{C}_q)_{i'j'} \\
     \end{aligned}
\end{equation}
Here $\mathbf{C}_q=\mathbf{C}_q'$ is the non-renormalized single 3JJQ capacitance matrix. The elements $(C_q)_{ii}$ are given by the sum of all the capacitances connected to the node $i$ when uncoupled, and the elements $-(C_q)_{ij}$ are given by the sum of all capacitances connecting nodes $i$ and $j$.The $\gamma_{ij'}$ are the coupling parameters in~\eqref{eq:capintL} which are $0$ if there's no capacitive coupling involving the corresponding nodes. And $\mathbf{C}_c$ is the mutual capacitance which accounts for the capacitive couplings between nodes
\begin{equation}
(\mathbf{C}_c)_{ij'}=\gamma_{ij'}C.
\end{equation}
This way, we find the renormalized inverse capacitance matrix of the qubits, $\widetilde{\mathbf{C}}_{q}^{-1}$ and $\widetilde{\mathbf{C}}_{q}^{'-1}$, and the inverse mutual capacitance matrix, $\mathbf{C}_c^{-1}$, by performing the inversion of the full capacitance matrix of the system:
\begin{equation}
    \mathbf{C}^{-1}=
        \begin{pmatrix}
        \widetilde{\mathbf{C}}_q^{-1} & \mathbf{C}_c^{-1} \\
        (\mathbf{C}_c^{-1})^T & \widetilde{\mathbf{C}}_q^{'-1}
        \end{pmatrix}.
\end{equation}

\clearpage

\section*{\label{sec:appendixB} Supplementary Material B: Complementary graphs for the different configurations and parameters studies}

\begin{figure*}[h!]
    \centering
    \includegraphics[width=\linewidth]{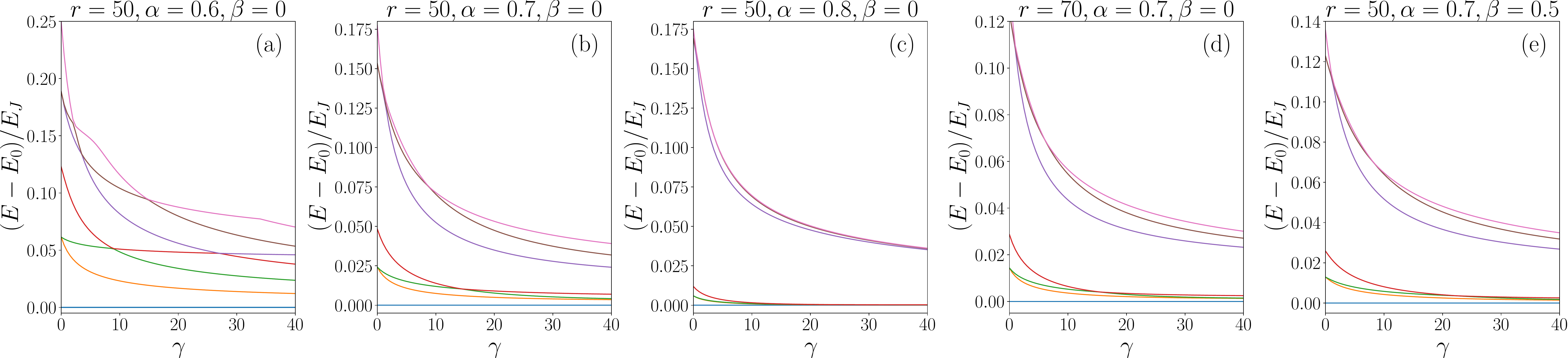}
    \caption{Low energy subspace as a function of $\gamma$ for the reference circuit, two 3JJQs with ground in $\phi_0$($\phi_0'$) coupled through a capacitor connecting nodes $\phi_2-\phi_1'$, and different qubit parameters: (a) $r=50$, $\alpha=0.6$, $\beta=0$, (b) $r=50$, $\alpha=0.7$, $\beta=0$, (c) $r=50$, $\alpha=0.8$, $\beta=0$, (d) $r=70$, $\alpha=0.7$, $\beta=0$, (e) $r=50$, $\alpha=0.7$, $\beta=0.5$.}
    \label{f. capacitive coupling multiple parameters}
\end{figure*}

In this supplementary material, we introduce some complementary graphs and explanations to provide further support and evidence for the results presented in the letter. In Fig.~\ref{f. capacitive coupling multiple parameters}, we show the energy spectra obtained from the reference capacitive coupling configuration (Fig.1(a)) for different values of the qubit parameters. As shown in Fig.~\ref{f. capacitive coupling multiple parameters}(a) for small values of the parameter $\alpha$ the qubit's anharmonicity is not sufficiently large and hence there is no distinction between the qubit subspace (four lower energy levels) and the exited subspace. Increasing $\alpha$ results in an increase of the qubit's anharmonicity avoiding this problem but significantly reducing the magnitude of the interaction and qubit's gap, this is shown in Figs.~\ref{f. capacitive coupling multiple parameters} (b) and (c). The effect in the spectrum of increasing $\beta$ and $r$ is really similar to that of $\alpha$, as noted in the main text: the subspace of the coupled qubits gains definition while the interactions and the qubit's gap decrease (Fig.~\ref{f. capacitive coupling multiple parameters}(d) and (e)).  The results commented here motivate the election of an value of $\alpha$ between $0.6$ and $0.9$ for our study and are in concordance with the results presented in the letter.\\

\begin{figure*}[h!]
    \centering
    \includegraphics[width=\linewidth]{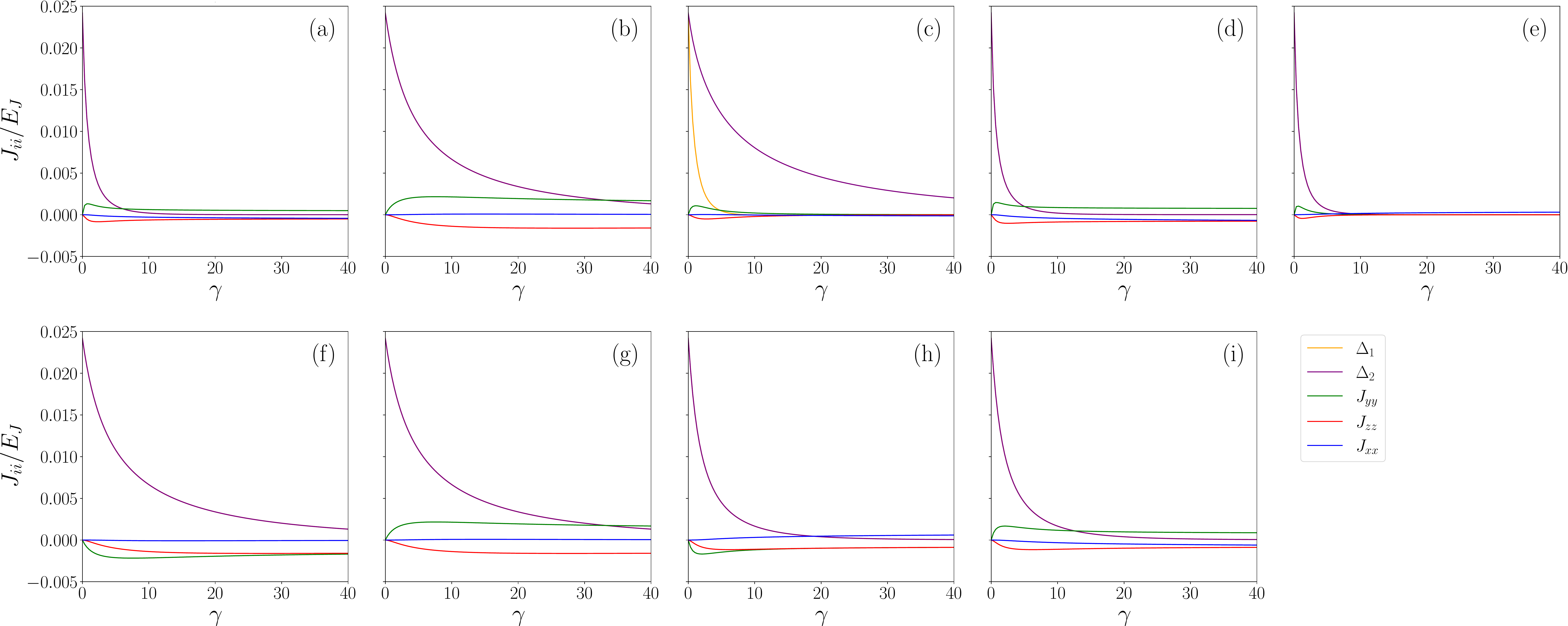}
    \caption{Effective Hamiltonian parameters as a function of $\gamma$ for different configurations of a circuit composed of two 3JJQs with $\alpha=0.7$ and $r=50$ coupled through a capacitor. (a-e) Ground in $\phi_1$($\phi_2'$): (a) coupled connecting $\phi_2-\phi_1'$, (b) coupled connecting coupled connecting $\phi_0-\phi_0'$, (c) coupled connecting $\phi_0-\phi_1'$, (d) coupled connecting $\phi_2-\phi_1'$ and $\phi_0-\phi_0'$, (e) coupled connecting $\phi_2-\phi_0'$ and $\phi_0-\phi_1'$.(f-i) Ground in $\phi_0$($\phi_0'$): (f) coupled connecting $\phi_1-\phi_1'$, (g) coupled connecting $\phi_1-\phi_2'$, (h) coupled connecting $\phi_1-\phi_1'$ and $\phi_2-\phi_2'$, (i) coupled connecting $\phi_1-\phi_2'$ and $\phi_2-\phi_1'$.}
    \label{f. capacitive coupling multiple configurations}
\end{figure*}

In Fig.~\ref{f. capacitive coupling multiple configurations} we present the coupling strengths and qubits gap for the different configurations of the circuit. First thing to check is that all the different circuit configurations result on a similar qualitative behavior of the constants when increasing $\gamma$: the qubits gap diminishes due to the renormalization, $J_{yy}$ grows until it reaches a maximum and starts to slowly decay and  $J_{zz}$ has a similar behaviour.\\

We consider two different ground configurations. In Fig.~\ref{f. capacitive coupling multiple configurations}(f-i) we plot the results for all configurations compound by two 3JJQs with ground in $\phi_0=\phi_0'=0$ coupled through one or two capacitors connecting different nodes. Here one must note that, since both qubits are identical and symmetric with respect to the $\alpha$-junction when their grounds are  in $\phi_0$ and $\phi_0'$, connecting nodes $\phi_1-\phi_1'$ is equivalent to the connecting nodes $\phi_2-\phi_2'$, whereas connecting nodes $\phi_1-\phi_2'$ is equivalent to connecting nodes $\phi_2-\phi_1'$. Comparing Fig.~\ref{f. capacitive coupling multiple configurations}(f) and (g) it can be extracted that changing the connection of the coupling capacitor from one corner of the qubit to the opposite corner does not change the magnitudes of the interactions, but it does change the sign of $J_{yy}$ and $J_{xx}$. The main reason for this is that changing the capacitors connections this way is equal to flipping the sign of the external flux threading one of the qubits, provoking a change in the sign of its charge and flux operators.  Fig.~\ref{f. capacitive coupling multiple configurations}(h) and (i) show that performing a double coupling of the qubits does not really increase the magnitude of the interaction but rather accelerates the renormalization of the gap and the appearance of the maximum coupling strenght for $J_yy$ and $J_zz$.\\

In Fig.~\ref{f. capacitive coupling multiple configurations}(a-e) we plot the results for all configurations composed of two 3JJQs with grounds in $\phi_1=\phi_2'=0$ coupled through one or two capacitors connecting different nodes. Here it is important to have in mind that the symmetries that applied in the previous case are no longer valid since the election of the ground breaks the qubits symmetry and equality, nevertheless, there are other equivalences that can be considered in this case given that the qubit's parameters are identical, and changing the qubit's ground from one upper corner to the other equates to changing the external flux's sign. For instance, configuration (c) is equivalent to connecting node $\phi_2-\phi_0'$ and any of the results that can be found by changing the ground connections from one upper corner to the other can be extracted from these graphics only by taking into account the change in the interactions sign. Broadly speaking, looking into Fig.~\ref{f. capacitive coupling multiple configurations}(a-c) we observe that connecting the coupling capacitor to the node opposite to the $\alpha$-junction or one of the upper corner nodes of the flux qubit produces different effects on its gap. This happens due to the way in which the capacitance matrix is modified when introducing the coupling, resulting in a quick renormalization of the gap for connections to the upper corner nodes, (a), and  an slow renormalization of the gap for connections to the lower node, (b). We can clearly see the difference in the circuit including a capacitor connecting an upper corner node with the lower node, (c). In this case we find a completely antisymmetric scheme where both qubits are renormalized differently. Focusing on Fig.~\ref{f. capacitive coupling multiple configurations}(d-e), one encounters again that doubling the coupling does not increase the magnitude of the coupling but rather accelerates the evolution of the coupling.\\

\begin{figure*}[h!]
    \centering
    \includegraphics[width=\linewidth]{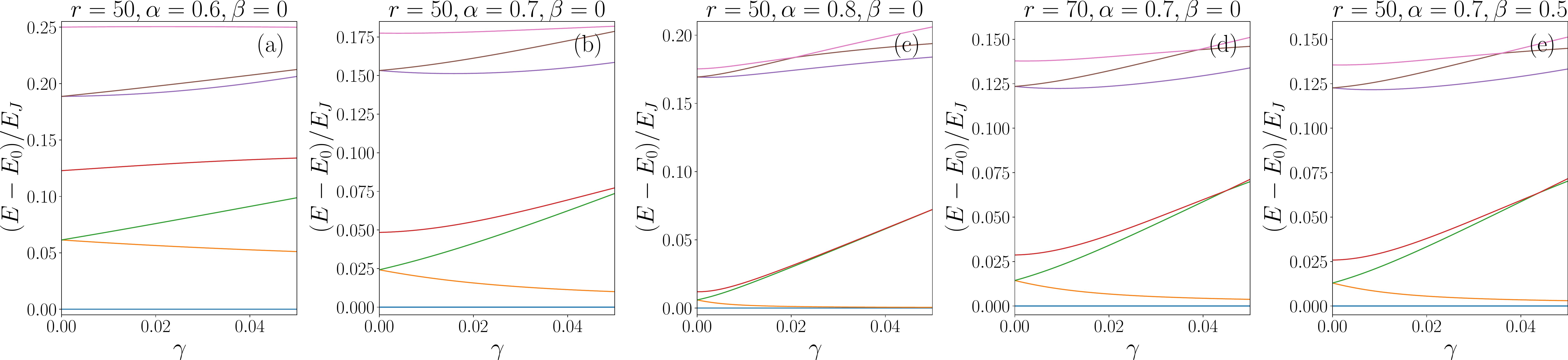}
    \caption{Low energy subspace as a function of $\gamma$ for the reference circuit, two 3JJQs with ground in $\phi_1$($\phi_2'$) coupled through a Josephson junction connecting nodes $\phi_2-\phi_1'$, and different qubit parameters: (a) $r=50$, $\alpha=0.6$, $\beta=0$, (b) $r=50$, $\alpha=0.7$, $\beta=0$, (c) $r=50$, $\alpha=0.8$, $\beta=0$, (d) $r=70$, $\alpha=0.7$, $\beta=0$, (e) $r=50$, $\alpha=0.7$, $\beta=0.5$.}
    \label{f. inductive coupling multiple parameters}
\end{figure*}

In Fig.~\ref{f. inductive coupling multiple parameters} we show the lower energy spectra of the reference Josephson junction coupling circuit for different values of the qubit parameters. For the range of $\gamma$ considered none of the values of the qubit parameters selected result in an extreme reduction of the coupling and qubit gaps or compromise the application of the SWT (the two subspaces remain separated), nevertheless, there are two important considerations that we have to make when interpreting these graphs. On one hand, it is shown in Fig.~\ref{f. inductive coupling multiple parameters}(a) that, even though no levels of the qubits subspace cross with levels in the high energy subspace, the two subspaces are not clearly differentiated in practice. The reduction of the qubit's anharmonicity for small values of $\alpha$ makes the distance between the two subspaces comparable to the qubits gaps. On the other hand, by looking into Fig.~\ref{f. inductive coupling multiple parameters}(b-e) we can check that increasing $\alpha$, $r$ or $\beta$ reduces the qubits gap (distances between levels 1-2 and 3-4)  but potentiates the growing of the coupling with $\gamma$ (distance between levels 2 and 3), providing with really large values of the coupling-gap ratio. However, this fast growth of the coupling may lead to crossing between the systems subspaces at relatively small values of $\gamma$.\\

\begin{figure*}[h!]
    \centering
    \includegraphics[width=\linewidth]{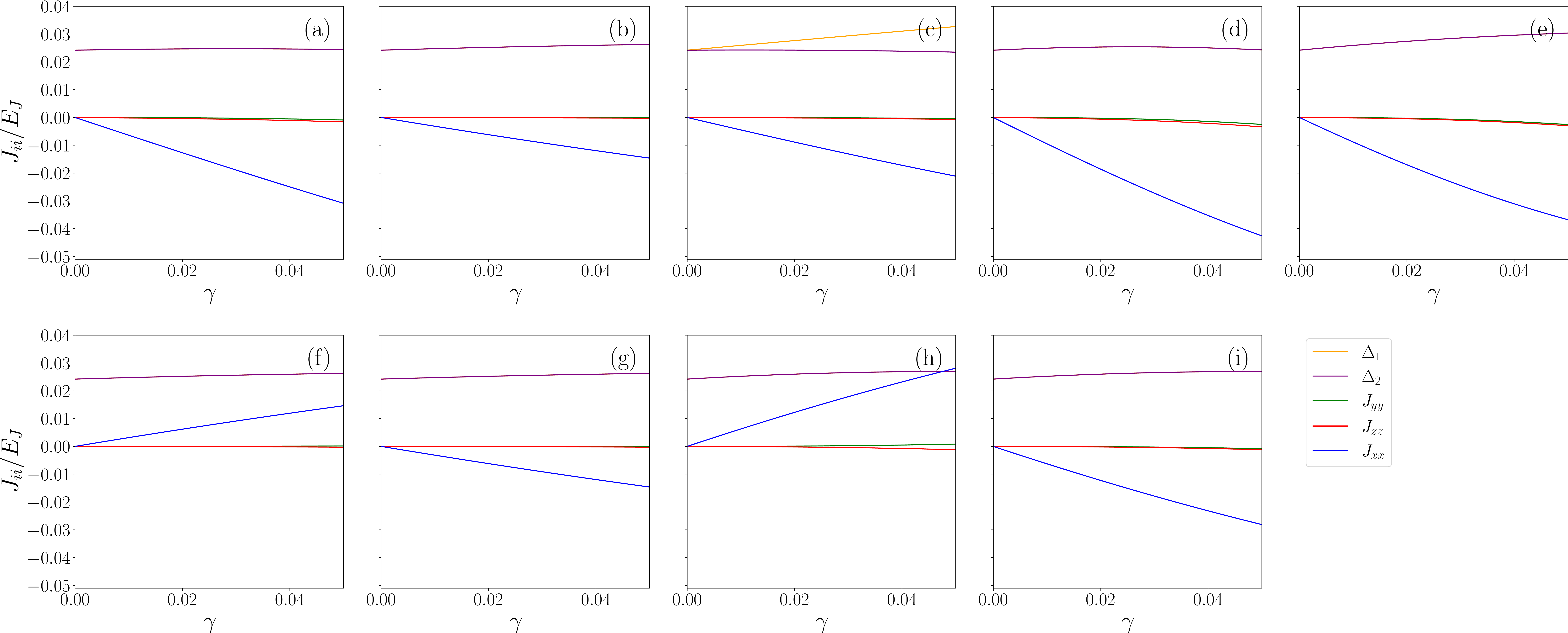}
    \caption{Effective Hamiltonian parameters as a function of $\gamma$ for different configurations of a circuit compound by two 3JJQs with $\alpha=0.7$ and $r=50$ coupled through a Josephson junction. (a-e) Ground in $\phi_1$($\phi_2'$): (a) coupled connecting $\phi_2-\phi_1'$, (b) coupled connecting coupled connecting $\phi_0-\phi_0'$, (c) coupled connecting $\phi_0-\phi_1'$, (d) coupled connecting $\phi_2-\phi_1'$ and $\phi_0-\phi_0'$, (e) coupled connecting $\phi_2-\phi_0'$ and $\phi_0-\phi_1'$.(f-i) Ground in $\phi_0$($\phi_0'$): (f) coupled connecting $\phi_1-\phi_1'$, (g) coupled connecting $\phi_1-\phi_2'$, (h) coupled connecting $\phi_1-\phi_1'$ and $\phi_2-\phi_2'$, (i) coupled connecting $\phi_1-\phi_2'$ and $\phi_2-\phi_1'$.}
    \label{f. inductive coupling multiple configurations}
\end{figure*}

In Fig.~\ref{f. inductive coupling multiple configurations} we present the coupling strengths and qubit gaps for different circuit configurations. As it can be extracted from the graphs the qualitative behavior of the coupling constants coincides for all circuit configurations: the qubit gaps remain practically unaltered, a predominant and growing $J_{zz}$ term appears and some residual $J_{zz}$ and $J_{yy}$ are found. The considerations that have to be made regarding these graphs do not differ from those made previously for the capacitive coupling, except for one important remark: in this case connecting two pairs of nodes with two Josephson junctions truly results in an advantage, since it results in larger coupling magnitudes (and not only larger $J/\Delta$ as in the double capacitive coupling) for small values of $\gamma$.\\

\end{document}